\documentclass[a4paper,gaps,physrev,showpacs,twocolumn,superscriptaddress]{revtex4-2}
\usepackage{amsmath,amssymb,mathrsfs}
\usepackage[pdftex]{graphicx}
\usepackage[dvipsnames]{xcolor}
\usepackage[normalem]{ulem}
\usepackage{physics}
\usepackage[british]{babel}
\usepackage{enumerate}

\newcommand\ContFracOp{%
  \operatornamewithlimits{%
    \mathchoice
     {\vcenter{\hbox{\huge $\mathcal{K}$}}}
     {\vcenter{\hbox{\Large $\mathcal{K}$}}}
     {\mathcal{K}}
     {\mathcal{K}}}}

\newcommand{\kk}{\mathbf{k}}
\newcommand{\bq}{\mathbf{q}}
\newcommand{\yy}{y}
\newcommand{\JJ}{\mathcal{J}}
\newcommand{\QUB}{School of Mathematics and Physics, Queen's University Belfast, Belfast BT7 1NN, Northern Ireland, United Kingdom}
\newcommand{\ETSF}{European Theoretical Spectroscopy Facility (ETSF)}
\newcommand{\IPPT}{Institute of Fundamental Technological Research, Polish Academy of Sciences, Adolfa Pawi\'nskiego 5b, 02-106 Warsaw, Poland}
\begin{document}
\bibliographystyle{unsrt}

\title{Generalized incommensurability: the role of anomalously strong spin-orbit coupling for the spin ordering in a quasi-2D system, FeSe}

\author{Piotr Chudzinski} 
\address{\QUB}
\address{\IPPT}
\author{Abyay Ghosh}
\address{\QUB}
\author{Myrta Gr\"uning}
\address{\QUB}
\address{\ETSF}

\date{\today}

\begin{abstract}
We study 2D spin and orbital systems, in a classical limit, in a regime where their coupling is so strong that orbital fluctuations are able to change sign of spin-exchange. Our aim is to understand how different phases in the orbital sector determine the ordering in the spin sector. The existence of intermediate vortex crystal (VC) phases, beside the canonical Kosterlitz and Thouless (KT) phases is now a well-established fact. Recently, we found [Phys. Rev. Research 6, 043154] that such a phase can explain phase diagram of FeSe. 
Motivated by this, here we develop a Renormalization Group (RG) method which can capture the formation of the VC phase. We achieve it by incorporating orbital vortex-vortex interactions through space-dependent incommensurability. Simultaneously, we incorporate the coupling with the spin sector for both short- and long-range interactions. We then derive a phase diagram including the effects of long-range interactions and spin-orbit coupling. The presence of the intermediate VC phase in the orbital sector changes the scaling of long-range spin-spin interactions, making them relevant under specific conditions. We thus find that the regular arrangement of orbital vortices can induce a long-range order in the 2D spin system.
\end{abstract}

\maketitle

\section{Introduction}
The Kosterlitz and Thouless (KT) phase transition~\cite{Kosterlitz_1973} offers a way to bypass Mermin-Wagner theorem in a two-dimensional (2D) system with a continuous symmetry and is one of the most inspiring discoveries in the field of correlated systems. In the KT transition, at a given finite temperature, topologically-bound vortex-antivortex pairs unbind into a free gas due to entropic effects. A half-century after its theoretical discovery, the two-dimensional systems have been realized and the KT transition confirmed experimentally in different settings: at the interface of two solids, in liquid crystal layers and even in cold atomic gasses. Further research explores how to amend the model to enrich its phase diagram beyond just two phases, enabling other, quasi-ordered phases~\cite{PhysRevB.103.024518,pnase2205845119,PhysRevE.102.032113}.
Particular focus has been on intermediate vortex crystal phases, that is on inhomogeneous phases where vortices are bound (e.g. by strings~\cite{PhysRevLett.55.541,PhysRevResearch.2.013081}) and arranged in large-scale patterns\cite{PhysRevE.99.022114,PhysRevE.99.062112,PhysRevE.97.052101,PhysRevB.85.094513,Lu_2012,PhysRevLett.106.067202,PhysRevB.84.224420,PhysRevLett.101.167202}.

Such intermediate vortex crystal phases  are possible thanks to vortex-vortex interactions, as postulated already in 1987 by Minhagen~\cite{PhysRevB.36.5620} and later proven both by mean-field arguments~\cite{gabay93} and by exact analytic~\cite{Campbell_1999} and numerical~\cite{coleman-numer} methods. 
An intriguing but not so much explored aspect is whether these intermediate vortex crystal phases are relevant in real materials. In our recent work~\cite{ghosh2024firstprinciples}, we identified FeSe---a quasi-2D, iron-based high-$T_c$ superconductor---as a candidate material where the vortex crystal phase may play a crucial role in defining the phase diagram. In the present work, we address the questions on \emph{(i) how the transition into the vortex crystal phase proceeds} and, for the case of FeSe, \emph{(ii) whether we can identify experimentally measurable quantities that can serve as a hallmark of this new physics}.   

In our ab initio study on FeSe~\cite{ghosh2024firstprinciples}, we have uncovered an extraordinary dependence of spin-exchange parameters on orbital composition at each Fe site. 
Based on that result, we proposed the following mechanism: a long-range order in the spin sector arises due to 2D vortex crystallization in the orbital sector. The loophole of this conjuncture is that the relation between the vortex crystal formation in the orbital sector and the long-range order in the spin sector has not been established so far. The renormalization group (RG) is perfectly suited to describe the relation between the spin and orbital sector as it can simultaneously capture both phenomena. Solving the RG equations would allow to observe the gradual change in the relevance of spin long-range interaction as the orientational order sets in the orbital sector at increasing characteristic distances. Through the solution of RG equations one could answer the highly non-trivial issue of whether vortex crystallization will be fast enough to ensure spin-long range order. Deriving such an RG procedure is the aim of the current study. 

However, before we focus on the issue of orbitally induced spin order, we need RG equations that capture the existence of vortex crystal on equal footing with the other two canonical KT phases. In fact, while the existence of the vortex crystal phase is a well-established\cite{Campbell_1999} fact, deriving the systematic RG procedure that would capture the new, generically inhomogeneous phase has turned out to be a hard task. An integro-differential formulation of RG has been put forward by Minhagen and collaborators\cite{Minnhagen-further1,Minnhagen-further2} when looking into a single mode of self-localizing bosonic fluctuations. The proposed procedure is quite complicated and lead to ambiguous results: the correct flow has to be chosen \emph{a posteriori} among several trajectories, some of which flowing backwards in RG time\footnote{in a way this resembled situation encountered in Van-der-Waals description of a gas-liquid transition}. These works were dedicated to a single mode of bosonic fluctuations which self-localize itself.

In our approach, we extend the problem by including two modes of bosonic fluctuations, i.e. spin and orbital. Moreover, the spin sector is permitted to have several minima, several orderings, in its potential energy landscape. These extensions both bring the model closer to the realistic FeSe platform and make the solution more straightforward and physically transparent despite the apparent higher level of complexity. Now, there is one unambiguous decoupling channel. We propose that a vortex crystal phase (in the orbital sector) can be captured by a spatially inhomogeneous landscape potential (due to the spin-sector), in effect a mean-field treatment of the vertex function. We call it \emph{a generalized incommensurability} that is emergent due to the orientational order present in a vortex crystal phase. Notice that this is distinct from the canonical understanding of incommensurability in low-dimensional systems, which is usually associated with missing carriers and prevents ordering---like in a doped Mott system. The generalized incommensurability is a spatially varying incommensurability that instead facilitates the developing of long-range order. This is a \emph{new mechanism of a phase transition in a low-dimensional system}, possible whenever two collective systems are coupled and one of them hosts the right number of vortices.

In what follows, we introduce the model and discuss its consequences for FeSe (Sec.~\ref{sec:state}). We then derive theoretical frameworks to describe the mobile vortex regime (Sec.~\ref{sec:vort-disord}) and the vortex crystallization (Sec.~\ref{sec:vortex-crystal}). The latter frameworks are used to obtain a RG flows, and we provide a proof that intermediate vortex crystal phase does exist in the orbital sector and can lead to long-range spin order (Sec.~\ref{sec:RG}). Finally, we identify measurable quantities to experimentally verify the proposed mechanism (Sec.~\ref{sec:meas}) and discuss the broader consequences of our results (Sec.~\ref{sec:disc}-\ref{sec:conc}).    

\section{Statement of the problem}\label{sec:state}

\subsection{The model}

In what follows, we consider a two-dimensional (2D) periodic system, with a substantial density of states at the Fermi energy---this is the case of metals or Mott insulators. For such systems, the spectral function of the material cannot be fully interpreted in terms of single (quasi)particle: instead a sizable part of the weight is moved to collective bosonic modes of orbital- and spin-degrees of freedom. These collective bosonic modes, and their coupling are the focus of this section and central to this work. In App.~\ref{app:A}, we show how these modes can be introduced formally through appropriate Hubbard-Stratanovich transformations.

\begin{figure}[h!]
 \centering
 \includegraphics [width=0.44\textwidth]{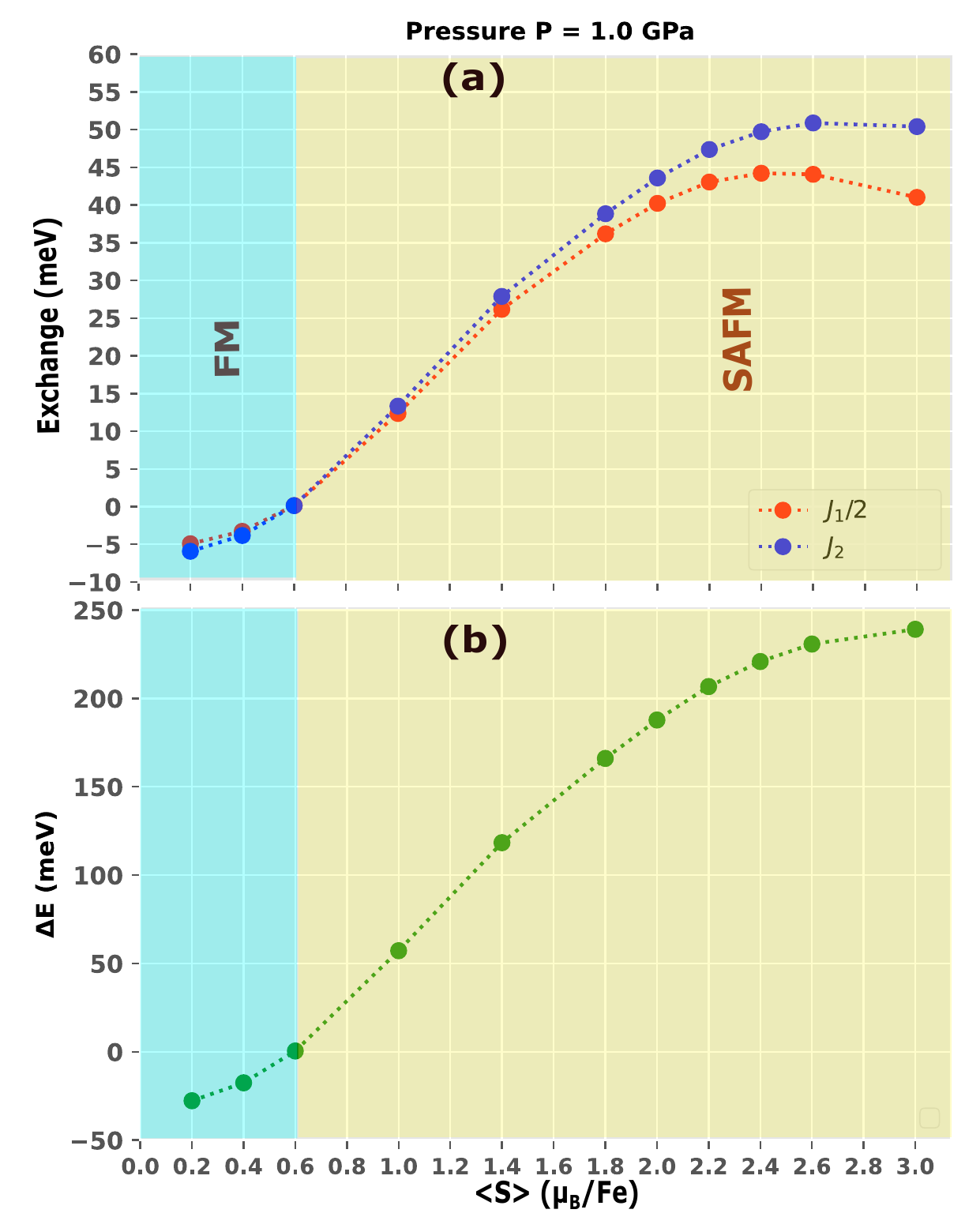}
 \caption{First-principles results of FeSe that inspired this work: Dependence of (a) nearest-neighbor $ J_1/2$ (red dots), and next nearest-neighbor $J_2$ (blue dots) spin exchange parameter  on average spin amplitude $\langle S\rangle$; (b) energy difference between ferromagnetic (FM) and striped antiferromagnetic (SAFM) phase for FeSe at 1 GPa pressure. Below 0.6 $\mu_B/$Fe the stable magnetic phase is FM, above the SAFM. All calculations were carried out within Density Functional Theory at the generalised gradient approximation level. $J_1,J_2$ have been obtained by fitting the results with a Heisenberg-like model Hamiltonian\footnote{We use the same approach and computational parameters as in Ref.~\cite{ghosh2024firstprinciples}}. }
 \label{fig:fugacities}
\end{figure}

To describe the dynamics of the orbital and spin components of the collective modes, we introduce the following Hamiltonians (in what follows, the subscript $L$ indicates the orbital component and $s$ the spin-component):
\begin{align}
  H_L[b_L] &= \notag\\ &\iint d^2r W_L (\nabla b_L(\vec{r}))^2 + \mathcal{J}_H (n_L(\vec{r}))^2, \label{eq:orb-def1} \\
  H_s[b_s] &= \notag\\ &\iint d^2r J[n_{L}] (\nabla b_s(\vec{r}))^2 + \mathcal{K}_{bi} [n_{L}] (n_s(\vec{r}))^2 \label{eq:spin-def1}.
\end{align}
Equation~\ref{eq:orb-def1} depends on $b_{L}$, the field operator that creates orbital fluctuation at a given point $\vec{r}=(x,y)$. The density $n_L$ of orbital fluctuations in terms of the corresponding boson creation/annihilation operators is,
\begin{equation}\label{eq:dens-flu}
n_L(\vec{r})= b_L^{\dag}(\vec{r})b_L(\vec{r}).       
\end{equation} 
The first term on the right hand-side of Eq.~\eqref{eq:orb-def1} corresponds to the kinetic part, with the material-dependent parameter $W_L$ being---in analogy with tight-binding ---the ``bandwidth of orbital excitations''. The second term describes the density-density interaction, which is in fact a four-fermion term known from Hund’s rules, and the Hund's $\mathcal{J}_H$ is a material-dependent parameter.
In analogy, Eq.~\ref{eq:spin-def1} depends on $b_{s}$, the field operator that creates spin fluctuation at a given point $\vec{r}=(x,y)$ and the corresponding density of spin fluctuations, $n_s$, defined in analogy with the orbital counterpart (Eq.~\ref{eq:dens-flu}), in terms of the corresponding boson creation/annihilation operators. Again, the first term on the right hand-side of Eq.~\ref{eq:spin-def1} corresponds to the kinetic part, tuned by the exchange parameter $J$, and the second term $\mathcal{K}_{bi}$ describes the density-density interaction. Here, we made the \emph{Ansatz} that the material-dependent parameters, $J$ and $\mathcal{K}_{bi}$, depends on the the density of orbital fluctuations, $n_L$. Crucially, this dependence is what provides the coupling between the orbital- and spin-degrees of freedom. This \emph{Ansatz} is justified by first-principles simulations of FeSe: the exchange parameters obtained from density-functional theory have a strong dependence on the orbital configuration of the Fe atom (determining the system magnetization), as seen in Fig.~\ref{fig:fugacities}(a).

We further assume \emph{classical} degrees of freedom, so their dynamics, in the continuum limit, can be described by the angles, $\theta_L, \theta_s$, between pairs of the orbital momenta and of spin vectors, $\vec L$ and $\vec S$. Correspondingly, we introduce the canonically conjugated scalar fields of phase $\phi_L$ and $\phi_S$. Then, the density of orbital- and  spin-fluctuations (Eq.~\ref{eq:dens-flu}) can be expressed as $n_\alpha(\vec{r})= \vec{e}\cdot \nabla \phi_\alpha(\vec{r})$ with $\alpha = L,s$ and where we project onto $\vec{e} = \mathcal{N}[1,(1\pm \zeta)]$ where $\zeta \ll 1 $ is the orientational order parameter defined later in the text and $\mathcal{N}$ is the appropriate normalization factor. 
Then, in the small angle limit 
, we can rewrite the Hamiltonians in Eqs.~\ref{eq:orb-def1}-\ref{eq:spin-def1} as
\begin{align}
  H^\text{TLL}_L[\theta_L, \phi_L] &= v_L \iint d^2r  K_L (\nabla \theta_L)^2 + \frac{1}{K_L} (\nabla \phi_L)^2 , \label{eq:orb-def2} \\
  H^\text{TLL}_s[\theta_s, \phi_s] &= v_s \iint d^2r  K_s (\nabla \theta_s)^2 + \frac{1}{K_s} (\nabla \phi_s)^2 , \label{eq:spin-def2} 
\end{align}
where we understood the spatial dependence of $\phi_s, \phi_L, \theta_s, \theta_L$ and the functional dependence of $K_s$ on $\phi_L$.  
Equations~\ref{eq:orb-def2}-\ref{eq:spin-def2} describe a Tomonaga-Luttinger liquid (TLL) for the orbital and spin components. In a classical 2D context, this is thus a model of two elastic membranes each defined by characteristic velocity and compressibility of acoustic waves $v_{\alpha}$ and $K_\alpha$, $\alpha = L,s$. The compressibility parameters $K_\alpha$ are non-trivially related to $W_L,J,\mathcal{J}_H$ and $\mathcal{K}$ in Eqs.~\ref{eq:orb-def1}-\ref{eq:spin-def1}. In particularly, when the interaction terms are neglected $K_{L}=W_L/T$ and $K_{s}=J/T$, where $T$ is the temperature.

While the small angle approximation is justified by the slow variations of $\theta_\alpha$ at low temperature, Eqs.~\ref{eq:orb-def2}-\ref{eq:spin-def2} are missing the essential low-energy physics in 2D systems from vortex-type excitations (see e.g. Ref.~\cite{Benfatto_12}). We thus include in the model large angles terms $\cos2\phi_\alpha$---topological excitations that induce $2\pi$ vortices in the canonically conjugated $\theta_\alpha$ field, 
\begin{equation} \label{eq:H_vort}
    H_\alpha^\text{vortex} =  \iint d^2r\, \yy_\alpha\cos2\phi_\alpha(\vec{r}),\quad \alpha = s,L.
\end{equation}
In Eq.~\eqref{eq:H_vort}, $\yy_\alpha$ is the density of vortices, called fugacity, with temperature dependent amplitude,
\begin{equation} \label{eq:TD_fugacity}
\yy_\alpha\propto\exp(-\mu_c^{(\alpha)}/T),    
\end{equation}
where $\mu_c^{(\alpha)}$ is the vortex core energy. For the spin-channel, an expression for the vortex core energy was given by Kosterliz and Thouless, 
$\mu_c^{(s)}= \epsilon_s J$ with $\epsilon_s = \pi^2/2$. 
This is the zeroth order approximation which includes the normalized energy cost $\epsilon_s$ of creating the vortex pattern of spins due to the exchange interaction $J$ in Eq.~\ref{eq:spin-def1} (in the sine-Gordon model this energy is also equal to the soliton mass of a given mode). By including screening~\cite{Benfatto_12} in vortex gas, one finds instead $\epsilon_s=3/\pi$. As discussed previously, in the model we proposed, $J$ has a functional dependence on the orbital fluctuations $n_L$, thus on $\phi_L$, namely $J[n_L]=J_0\cos\phi_L$ where $J_0$ is the bare amplitude of exchange. Then,
\begin{equation}\label{eq:fugacity_s}
\yy_s \propto \exp(-\beta\epsilon J_0\langle\cos\phi_L\rangle)    .
\end{equation}

Equations ~\ref{eq:orb-def2}-\ref{eq:spin-def2} together with Eq.\eqref{eq:H_vort} constitutes a standard description of classical 2D spin system in the continuous limit. It is known that the only phase transition it supports is the famous Kosterlitz-Thouless phase transition: the vortex unbinding into free gas. This is supported by 
Mermin-Wagner theorem, which states that when spin-spin interactions are short range there is no phase transition at any finite temperature for a classical 2D systems. Then, to introduce spin ordering and thus, a mechanism for a phase transition in the classical 2D spin system of Eq.~\ref{eq:spin-def2}, we need to introduce in addition the long-range (LR) interaction term,
\begin{equation}\label{eq:H_spinLR}
    H_s^\text{LR} = \iint d^2r \frac{\yy_\text{LR}[\phi_L]\cos\theta_s(\vec{r})}{\sqrt{(x^2 + y^2)^\gamma}}, 
\end{equation}
where $\yy_\text{LR}$ is the amplitude of the process that has been written as spin-tunnelling term\cite{BKT-long-range} (i.e. proportional to $\cos \theta_s$), and the parameter $\gamma$ sets the decay rate of the long-range spin-spin interaction. In order to induce the long-range order the exponent $\gamma$ must be equal or smaller than the dimensionality of the system, $\gamma \leq 2$.

We sum the vortex [Eq.~\eqref{eq:H_vort}] and long-range [Eq.~\eqref{eq:H_spinLR}] terms to Eqs. \eqref{eq:orb-def2}-\eqref{eq:spin-def2} and obtain our final model Hamiltonian $H_\text{col}$ for the coupled spin- and orbital-collective modes,
\begin{align}\label{eq:ham}
    H_\text{col} &= v_L \iint d^2r  K_L (\nabla \theta_L)^2 + \frac{1}{K_L} (\nabla\phi_L)^2 \notag \\
    &+ v_s \iint d^2r  K_s[\phi_L] (\nabla \theta_s)^2 + \frac{1}{K_s[\phi_L]} (\nabla \phi_s)^2 \notag \\
    &+  \iint d^2r\, \yy_L \cos2\phi_L + \yy_s [\phi_L] \cos2\phi_s \notag \\
    &+\iint d^2r \frac{\yy_\text{LR}[\phi_L]\cos\theta_s}{\sqrt{(x^2 + y^2)^\gamma}},
\end{align}
where we understood the spatial dependence of $\theta_\alpha,\phi_\alpha$, $\alpha = s,L$. 

Note that, since we are considering nonlinear cosine terms in the Hamiltonian, the spin- and orbital-fluctuations have both a plane-wave and a soliton-like---carrying a localized angular momentum---components. In what follows, we address the soliton-like bosonic fluctuations of spin origin as solitons ad of orbital origin as orbitons. When the boson-boson coupling is of a local displacement type, which should hold for small displacements, then $V_{LS}(q)\sim q$. Thus interaction with small momenta exchange, i.e. interaction of densities $\nabla\phi_s \nabla\phi_{L}$, is of a minor importance. Anyway, including them would only re-diagonalize the bosonic modes, leading to hybrid spin-orbit modes. It is well known how this can be accommodated, but for now we neglect it. In this work we shall focus on local excitations with large momentum exchanged, that is coupling between topological excitations, solitonic waves.

Some observations on how the model Hamiltonian relates to the target material, FeSe. first, the long-range term proportional to $y_\text{LR}$ in the Hamiltonian is particularly relevant for FeSe, for which several works~\cite{Glasbrenner2015,Wang2016,cite-key6,Fernandes2022,Sun2016} have emphasized the importance of the long-range nature of spin-exchange interactions. In fact,  from ab-initio results,  we found~\cite{ghosh2024firstprinciples} that spin fluctuations dispersion strongly depend on the orbital degree of freedom, and that orbital fluctuations have finite dispersion. This implies that spin-spin interactions can be long-range and mediated by orbital degree of freedom. Then, the characteristic decay exponent of spin-spin interactions is determined by the orbiton propagator:
$$
y_{LR}\propto \langle b^{\dag}(r)b(r')\rangle=\exp(-\langle\theta(r)\theta(r')\rangle).
$$
Second, it is experimentally well-established that only the high-pressure phase of FeSe is magnetically ordered~\cite{Sun2016}. From ab-initio results~\cite{ghosh2024firstprinciples}, we observed that the principal difference between low- and high-pressure phases is the twofold reduction of $W_L$ defined in Eq.~\eqref{eq:orb-def1}. Such reduction substantially changes the fugacity $\yy_L$. In turn, when the orbital content changes, then the fugacity of the spin sector $y_s$ changes because $J$ depends on the orbital content, see Fig.\ref{fig:fugacities}(a).

\subsection{Expected results: qualitative description}\label{ssec:RG-qualit}

\paragraph{RG equations: General idea} The Hamiltonian [Eq.~\eqref{eq:ham}] for the spin and orbital degrees of freedom consist, for both sectors, of density wave fluctuations $(\nabla\phi)^2$ and a cosine term describing topological vortex excitations. The first term is solvable by Gaussian averages while for the second term we use the RG method, which is a well established tool in this context (see App.~\ref{app:B1} for details). Within the RG, short range fluctuations, with ``$<$" index,  are gradually averaged out to produce an effective theory at macroscopic scales. For instance, the compressibility (we understood the L/s subscripts) changes like:
\begin{multline}\label{eq:RG-basic}
    \delta K[\Lambda]=\\
    y^2 \int_{\Lambda}^{\Lambda'} dr_{1}dr_{2}\cos(\theta^{>}(r_1)-\theta^{>}(r_2))\langle\exp(\theta^{<}(r_1)-\theta^{<}(r_2))\rangle_{\Lambda}\\
    =y^2\int_{\Lambda}^{\Lambda'} dr r^3 \langle\theta^{<}(r)\theta^{<}(0)\rangle_{\Lambda} J_n(\Lambda r),
\end{multline}
where $r\equiv (r_1-r_2)$, the $J_n$ is the Bessel function of the $n$-th  order (the choice of $n$ depends on whether we intend to compute monopole, dipole, quadrupole\dots correlation on the plane) and the key quantity that contains the physics of the problem is $\langle\theta^{<}(r)\theta^{<}(0)\rangle$. Note that for a 2D classical system, contrary to 1D quantum system, the RG flow does not probe decreasing energy shells. Instead, for a given temperature $T_0$, one increases the characteristic length on 2D plane from $\Lambda=a_0\exp(l)$ to $\Lambda'=a_0\exp(l+dl)$, where $a_0$ is the crystal lattice constant (the UV cut-off) and $l$ is the running variable in the RG equations. By this process, one examines if the effective density of \emph{free} vortices increases or decreases. If it increases, $y_i$ is relevant and the effective theory is ultimately given by vortex gas. Instead, if the effective density decreases, $y_i$ is irrelevant and the  effective theory is defined only by density waves because any remaining vortices are bound. This is a standard KT picture. What is less known, although proposed more than three decades ago~\cite{Minhagen-rmp}, is that an intermediate phase of vortex crystal with a constant $y_i$ is also possible. In this phase vortex-antivortex excitations still form bound pairs, but the density of these ``dipoles" is so large that their interactions matter and a crystal of dipoles is formed. We shall investigate the effects induced by the presence of this phase in \emph{one} of the sectors of the theory, in the orbital sector to be specific.  

The crucial microscopic mechanism of the system under consideration, that lays foundation for our field-theory, can be read out from Fig.~\ref{fig:fugacities}(a): varying the orbital occupancies of $d-$orbitals $n_{di}=\langle d_i^{\dag}d_i \rangle$, which is equivalent to change the $n_L$ or $\theta_L$, away from its optimal value, i.e. reducing $\langle S \rangle$ from high to low spin value, lowers the spin-exchange $J$ parameter, thus reducing the core energy $\mu_c$ for the spin vortex. Whenever an orbital vortex-antivortex pair appears in the 2D plane we shall have approximately an ellipse with eccentricity $\zeta$ and inside this ellipse the parameters of the spin system are different than outside. The spin-orbit coupling is thus quite unusual: the appearance of an orbiton increases the probability of spin vortex appearance in the orbiton's core. The presence of the spin vortex, $\propto \exp(i\phi_s)$, affects the long-range spin-spin interactions that are proportional to the canonically conjugate $\theta_s$ field.

\paragraph{Initial flow.} The RG flow starts at the shortest, atomistic scales with certain initial values for $g_{s,L}$ and $K_{s,L}$. These values depend on the temperature of the system. We a choose 
temperature $T_i\in (80K, 200K)$, as this is the temperature range of the parent state of the nematic/magnetic transitions in FeSe, the material of our interest\footnote{the parent state for superconducting transition is the emergent state, where some form of ordering is already present in the system, this is outside the main scope of this study.}. At this temperature, the spin system is well below the KT temperature, $T_{i}\ll T^{0}_{KT}\propto J_s$ (see Fig.~\ref{fig:fugacities}(a). This is because we take the value of $J_s\approx 50$meV , i.e. the value found (Fig.1a) for the high-spin (HS) striped antiferromagnetic (SAFM) phase. On the other hand, when $T=T_i$ the orbital system is in the regime close to the KT transition with rather large fugacity of orbital vortices, this is because of large Hund's $\JJ_H$ that lowers the value of the effective $K_L$.

\begin{figure}[h!]
 \centering
 \includegraphics [width=0.44\textwidth]{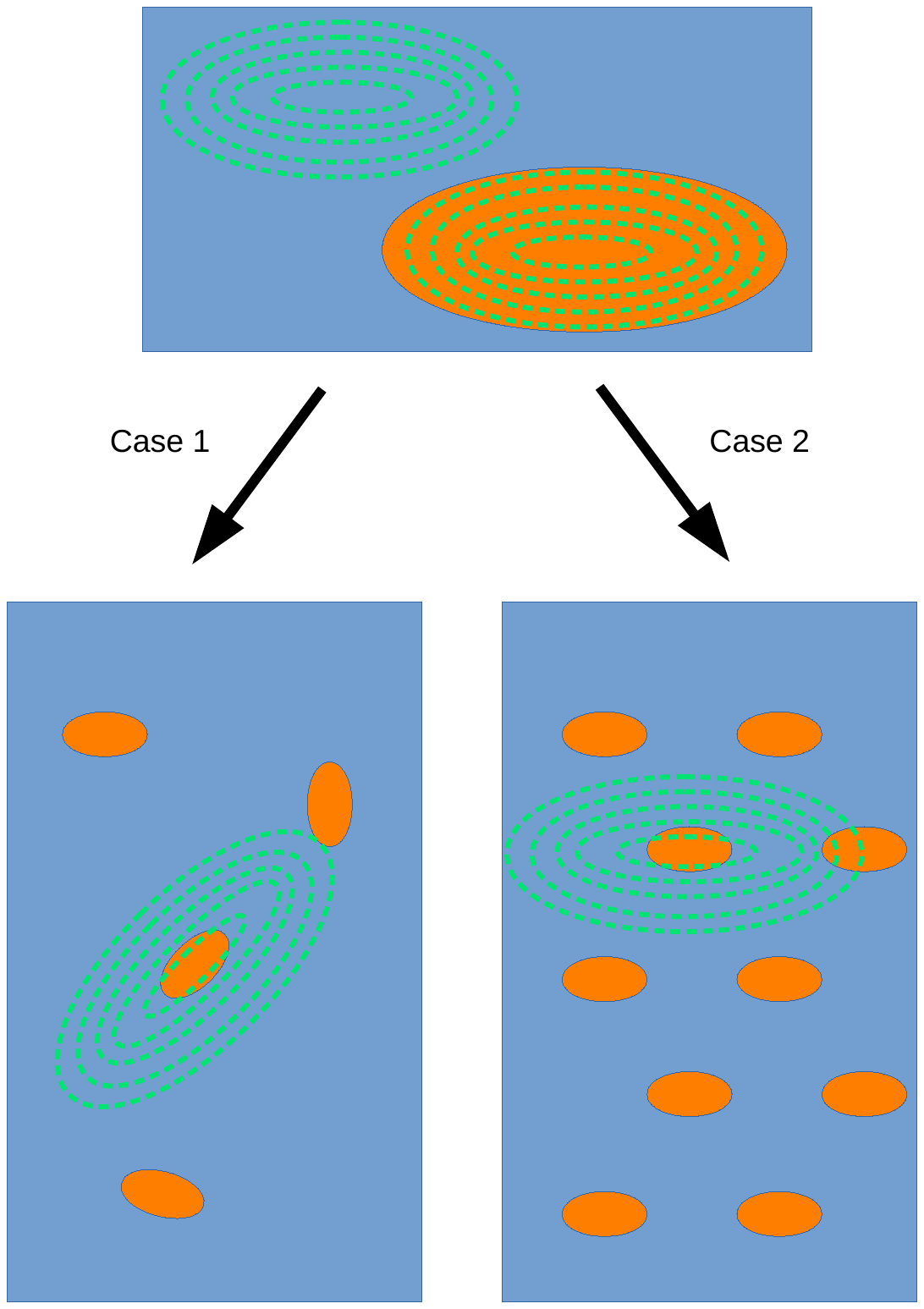}
 \caption{Schematic view of RG procedure. The characteristic size of the system (green ellipses) is increased to investigate the behavior of the system at increasing length-scales. Top panel (2D plane zoomed in): the initial flow, when the size is smaller than the low-spin (LS) zones. Bottom panels (2D plane zoomed out): two variants of the flow for larger length-scales when the densities of the LS zones are small (left panel) or large (right panel). The blue background indicates the high-spin (HS) zones, while the orange ellipses are the LS zones. The LS is equivalent to large distortion of $\theta_L$ value, that is the orbital-vortex core. In the left panel, the ellipses orientation is random thus the orientational parameter $\zeta=0$, while in the right panel $\zeta\neq 0$ has the same value in entire area shown.}
 \label{fig:RG-schem}
\end{figure}

The RG flow is schematically illustrated in Fig.~\ref{fig:RG-schem}. In the beginning of RG, that is close to the UV (short distance) limit, where one consider distances of order of few lattice sites, the RG flow proceeds along the standard KT trajectories. Because $J$ (and thus vortex core energy) is large there are not too many spin-vortices in the uniform, energy-favored SAFM (HS) phase, the long-range coupling $y_\text{LR}$ is expected to grow. On the contrary, in the orbital sector there is a substantial, relevant number of orbital vortices. Inside orbital vortices we have a spin ferromagnatic (FM) phase, hence AFM waves are prohibited from entering inside. Furthermore due to smallness of $J$ locally, the spin vortices may be present inside these orbital-vortices, these are spin-vortex excitations within FM phase.

\paragraph{Two possible outcomes} The second stage begins when the RG procedure (the flow) reaches the scale of orbital vortex size.  The flow of FM zones is now complete because we reached maximal size of these regions. The flow of SAFM and orbital sector continues, but with a different boundary conditions and initial conditions set by the 1st stage of the flow. A more complicated set of RG equations is now needed, we derive them in Sec.IIIB-C and write the entire system of equations explicitly in Sec.~\ref{sec:RG}. There are two possibilities for the flow:
\begin{enumerate}
\item Orbitons are diluted. Then, orbitons act as random impurities in an otherwise homogeneous medium. Orbitons and spinons are localized. If orbitons mediate long-range spin-spin exchange interactions, then this long-range part $y_\text{LR}$ is suppressed. This case (shown in the left panel in Fig.~\ref{fig:RG-schem}) is tackled in Sec.~\ref{sec:vort-disord}.
.\item The distances between orbitons are only one order of magnitude larger than the orbital vortex size. Then the system enters an intermediate range of RG with inhomogeneous, gradually changing local chemical potentials. Within this phase orbitons' fugacity can grow and then saturate, thus leading to substantial $\zeta$ ( the anisotropy which corresponds to the orbital vortex crystal parameter) and smaller exponent $\gamma$ [Eq.~\eqref{eq:H_spinLR}]. Then $y_\text{LR}$ can grow. This case (shown in the right panel in Fig.~\ref{fig:RG-schem}) is tackled in Sec.~\ref{sec:vortex-crystal}.
\end{enumerate}  

When case 1 is realized the last stage of the flow is spin-system with short range interactions (thus no order, or at best nano-domains of long-range order\footnote{LRO flow in the initial stage and short-range order in latter stages would imply nano-domains}) and orbital system's flow stops at a small fugacity of orbital vortices.

When case 2 is realized the last stage of RG leads spin system towards long range order. The two stage mechanism is as follows: when orbital vortex lattice form the background potential experienced by the $\phi_L$ field is oscillating, the fugacity inside crystal phase stays in average constant. We develop formalism to capture this in the beginning of Sec.~\ref{sec:vortex-crystal}. If the spin system is coupled with orbital system and the orbital system develops crystal vortex phase then spin-spin interaction can be mediated through the vortex lattice. We take $\zeta$ to be the amplitude of orbital vortex crystal, finite $\zeta$ means that the ellipses are oriented in the same direction over a finite area, however the distortion can be primarily either along x- or y-directions which leaves us with 2D Ising variable, known\cite{Wu-Ising-corr} to be decaying as $(1/r^{1/8})^2=1/r^{1/4}$. Thus the effective decay exponent of $y_\text{LR}$ can be approximated as $\tilde{\gamma}=3(1-\zeta)+\zeta/8$ (see Appendix B2). The system indeed moves towards a spin long-range order (LRO) phase, provided that we can prove the existence of a finite $\zeta$.

\section{Explicit spin-orbit coupling}

In order to solve the problem we need to express the implicit spin-orbit coupling in  the last two terms of Eq.\eqref{eq:ham} in the explicit form, i.e. to write explicit expressions for the following terms in the Hamiltonian:
\begin{align}
H_{s}^\text{vortex}[n_{L}(x,y)]&\rightarrow H_{sv(+L)},\\
H^{LR}_s[n_{L}(x,y)]&\rightarrow H_{LR(+L)},\\
H_{LS}[\vec{s}(x,y)]&\rightarrow H_{LS(+s)}.
\end{align}

where the terms in the right column contain an explicit functional dependence on $\phi_L$ of the spin compressibility $K_s$ and fugacities $\yy_s$ and $\yy_\text{LR}$ in Eq.~\ref{eq:ham}. The explicit form of the coupling between the orbital- and spin-fluctuations is needed for developing the RG, a necessary step to understand the phase diagram of the problem.

The simplest solution to the problem when there is a dependence on $J[n_L]$ is through a weak-coupling approach, that is with the two bosons weakly mixing. This is however not the regime we work on: in our case, $\Delta J[\Delta n_L]$ is of the same order of magnitude as $J[n_L=0]$ itself (see Fig.~\ref{fig:fugacities}). Then, the system can reach the regime with either \textbf{A)} $J\rightarrow 0$ so that a substantial spin vortex fugacity is present or even with \textbf{B)} $J$ changing sign and thus with the spin-character of the ground state $|\Psi_s\rangle$ being affected. In the latter case, to account for the changing spin-character of the ground state, we define a range $r_\text{FM}$ such that when $r$ belongs to this range it fulfills the following condition:
$$r \in r_\text{FM} \Rightarrow n_L(r)>n_{L0}: J[n_{L}>n_{L0}]<0,$$
where $n_{L0}$ is the amplitude of orbital fluctuation for which $J$ changes sign. 
In what follows, we investigate how the presence of finite $r_\text{FM}$ ranges affect the orbital degree of freedom. 

\subsection{UV limit, initial parameters for the second stage of the RG flow}\label{sec:UVmicro}

Let's focus on a situation with strong spin-orbit coupling, that is a large $\Delta n_L$. We can employ an analogue of \emph{Migdal} theorem (since the velocities of the two modes are quite different) and consider two velocities separately. A change in orbital density causes large change in the parameters of spin Hamiltonian. We thus reach case \textbf{B} with a number of long-living quasi-static FM cores distributed in the system. The RG procedure below assumes a given ground state $|\Psi_s\rangle$ around which $\theta_s(r)$ fluctuations are possible. Here, we have topologically different fluctuations in the sense that for a given range $r_\text{FM}$, as defined above, the ground state around which fluctuations are defined is different:
\begin{equation}
   |\Psi^\text{FM}_s(r\in r_\text{FM})\rangle \neq |\Psi^\text{AFM}_s(r\not\in r_\text{FM})\rangle ,
\end{equation}
This \emph{local} difference can be distinguished only at the microscopic level and thus it needs to be accounted for in the UV limit, in the initial parameters of the RG flow.

In the strong-coupling atomistic limit---the UV limit of our theory, one predicts that any finite value of $\JJ_H$ results in what is called the Hund's super-exchange. Namely, if the electrons on two neighboring sites have the same spin orientation they can swap through hybridization and due to the indistinguishability principle gain energy from a non-local exchange. This process, through a Hartree diagram, is also associated with the spin-orbital dependent hybridization\cite{Held-Jpropt} (see also App.~\ref{app:B}), $t_{LS}\propto \JJ_H$. Then, an additional term should be added to single-particle Hamiltonian (as written in App.~\ref{app:A}):
\begin{equation}
    H_{1p}^\text{add}=\sum_{i,j}\sum_{m,n,\sigma_i,\sigma_j}t_{LS}[\sigma_i,\sigma_j,m,n]c^{\dag}_{in\sigma_i}c^{}_{jm\sigma_j},
\end{equation}
where $m,n$ are atomistic orbitals and $\sigma$ indicates local expectation value of spin, the hybridization may depend in general on both these degrees of freedom.

If the neighboring sites spins are anti-parallel, then there is no energy gain. This implies the existence of an additional potential experienced by fermions that depends on spin orientation of sites coupled by the hybridization:
$$
V_{LS}(r,r')=\frac{t_{LS}}{2}|S|(\sigma(r)\cdot\sigma(r')+|\sigma|^2), 
$$
where the pre-factor has been computed in a similar manner as for the super-exchange (see App.~\ref{app:B} for details). Indeed, the existence of such an additional term can be also inferred from first-principles results, Fig.~\ref{fig:fugacities}(b), where the energy of orbital-fluctuation changes, depends on the $|\Psi^{i}_s\rangle$ phase.

 The $ H_{1p}^\text{add}$ term distinguishes between various magnetic phases that are coupled with orbital degree of freedom: 
 for SAFM it is $t_{LS}/2$, while for FM it reaches the maximum value, $t_{LS}$. 
Thus the energy levels of FM and AFM phases split and fluctuations between them are blocked. Since the $\JJ_H$-driven mechanism is the only source of a finite amplitude $t_{LS}$ considered here then in what follows we address its variation as $\delta t(r)$. Following the above given extra potential $V_{LS}$ we deduce that:
$$
\delta V_{LS}(x,y)\propto \delta t(x,y),
$$
that is, moving between different magnetic phases changes the local configuration of the orbital field.

\begin{figure}[h!]
 \centering
 \includegraphics [width=0.6\textwidth]{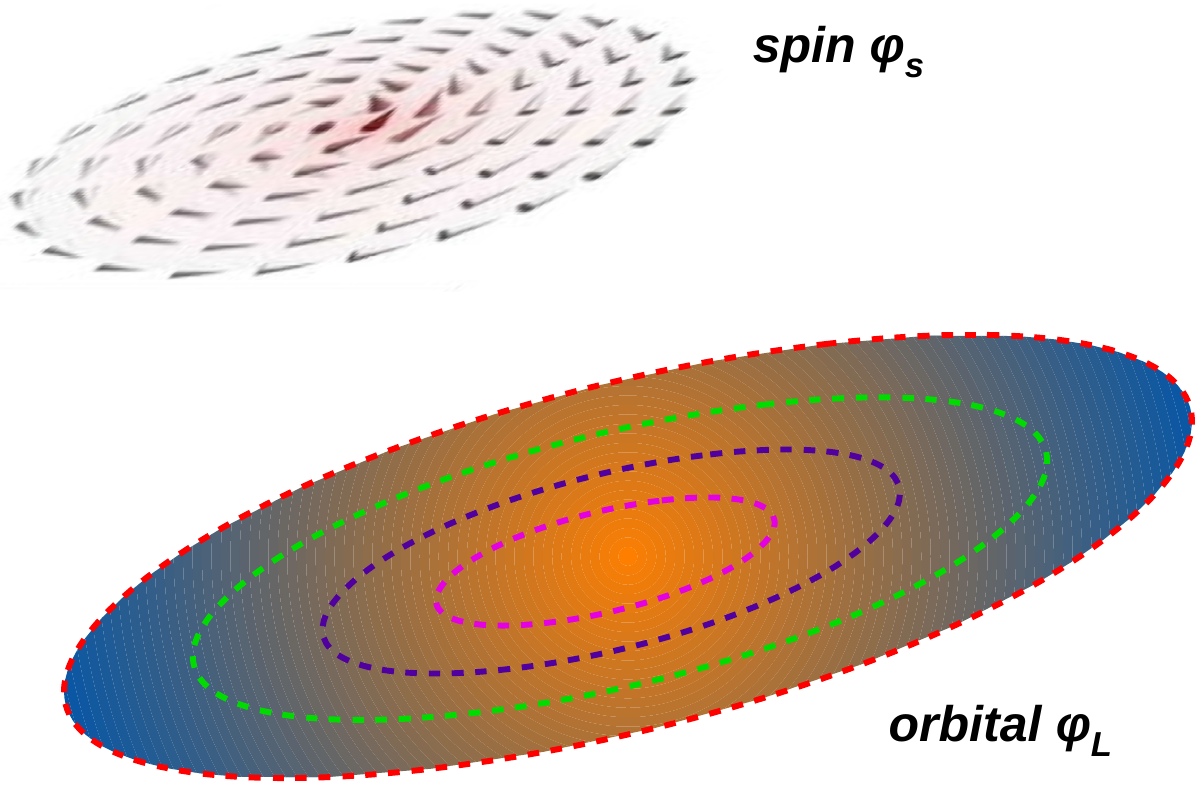}
 \caption{Schematic view of two overlapping vortexes in spin (top) and orbital (bottom) sectors. For the orbital $\phi_L$ the value is encoded in color: from Fig.~\ref{fig:fugacities} we deduce that the orange area inside red-circle is in $|\Psi^\text{FM}\rangle$ state, while in between red and purple circles $J \rightarrow 0$ and thus also $\nu_s \rightarrow 0$. In other words the coordinates of orange area inside red circle fulfill the condition to be $r_\text{FM}$}
 \label{fig:sLvort}
\end{figure}

We now move to vortex physics. With several possible ordering phases predicted in the spin sector, it is easier to begin with the orbital sector. It should be noted that usually too large (local) density of spin vortices is prohibited due to high energy cost: in an AFM environment some spins would need to be nearly parallel, which in a typical case is costly. However, the presence of an FM core changes the situation: the parallel arrangement of spins will actually be preferable in such region. Moreover, in the area where $J$ changes sign the spin-vortex core-energy goes to zero. Thus, spin-vortices are attracted towards the orbital-vortex as shown in Fig.~\ref{fig:sLvort}.

To substantiate this conjecture we follow this self-consistent procedure:
\begin{enumerate}
    \item we assume that in a strong coupling regime there is a number of spin vortices in the system. They form a quasi-static configuration because $|\Psi_s^\text{FM}\rangle$ states are topologically protected at microscopic, UV level;
    \item we explore how the existence of these spin-vortices affects the orbital system: in a dilute case (Sec.~\ref{sec:vort-disord}) we can solve correlation functions exactly, while in a dense regime (Sec.~\ref{sec:vortex-crystal}) we derive the $H_{LS(+s)}$ Hamiltonian and its renormalization;
    \item based on that, we derive the effective $H_{sv(+L)}$ and $H_{LR(+L)}$, study their impact on the RG and confirm the conjecture from step 1.  
\end{enumerate}

In step 1, we compute an average of any measurable $\hat{O}_L$ in the orbital sector in the presence of the spin-sector and the spin-orbital coupling: 
\begin{equation}
\langle \hat{O}_L \rangle = \frac{1}{Z}\int d\phi_i O_L[\phi_i]\exp(-S_L[\phi_i]),    
\end{equation}
where $S_L$ is the corresponding action. We assume that orbital field is perturbed by an averaged quasi-static vector field profile $\phi_L(x,y)\rightarrow\phi_L(x,y)+\delta t(x,y)\eta(x,y)$---a correction that is coming from the spin degrees of freedom in an (initial) mean-field approximation. To find the optimal solution for the energy expectation value we extremize the action $S_{L}(x,y)$: following the textbook procedure, it is equivalent to the solution of the Euler-Lagrange equations with
a variation:
\begin{equation}
    \delta t(x,y) \int_{C(x,y)}d^2r \frac{1}{K_L}v_L\nabla\phi_L(x,y)\nabla\eta(x,y)\rightarrow 0,
\end{equation}
where, as usual, through integration by parts the term $\sim \eta(x,y)$ appears. Assuming that boundary contribution on the contour $C(x,y)$ is zero, one finds that the Euler-Lagrange equation for the orbital sector stays unaffected, only the elastic part is present. This is the case for the density waves profiles, i.e. $\eta(x,y)$ is a plane wave with amplitude going to zero. The situation changes when there is a topological excitation, a spin-vortex texture enclosed inside the loop (this is the case \textbf{A} with potentially large vortex fugacity), since then, by the Stokes theorem, $\curl~\vec{\eta}(x,y)\neq 0$ and the boundary integral produces a finite value. An additional boundary contribution to Lagrangian $\propto \nabla\phi_L$ arises. This adds a term $\delta t(x,y) \nabla \phi_L$ to the TLL Hamiltonian of the orbital sector. This extra term gradually increases as the contour shrinks towards the center of the vortex. Two situations are possible: i) the field $\delta t(x,y)$ consists out of ``rare incursions" which can be treated as Dirac deltas; ii) the field is dense thus the internal structure of $\delta t(x,y)$ matters. These two cases correspond to cases 1-2 in the qualitative description in Sec.\ref{ssec:RG-qualit}c.




\subsection{Mobile vortices: disordered case}\label{sec:vort-disord}

In the case i), the extra term can be treated following the same lines as for the standard forward-disorder term in the TLL: 
we introduce a phase shift:
$$\phi_{L}(r)\rightarrow \phi_{L}(r)\int^r dr' \delta t(r') $$
that allows to gauge out the additional spatially varying field $\delta t(r')$. Upon this shift, we see that all correlation functions that contain the $\phi_L(r)$ field decay exponentially because of an extra pre-factor 
$$\langle\phi_L(r')\phi_L(r)\rangle\propto\\$$
$$\exp(\int dr\int dr' \delta t D(r-r'))=\exp(-\delta t |r-r'|)$$
where we assumed that the $\delta t(r)$ field is uncorrelated, random and thus $\langle\delta t(r)\delta t(r')\rangle=\delta t D(r-r')$ with $\delta t $ amplitude of hopping fluctuations and $D(r-r')$ the Dirac delta distribution.

We now look how the spin sector is affected by this weak localization of the orbital $\phi_L$ field. We have two nonlinear interaction terms in the spin sector: the spin-vortex term $H_{sv}$ and the long-range interaction term $H_\text{LR}$, each should be treated separately. For $H_{sv}$, we know that in FeSe there is a strong dependence $J(\delta \langle b_{L}^{\dag}(x)b_{L}(x)\rangle)$, to be precise, $$J_{s}\propto \langle\cos[\phi_L(x,y)-\phi_L(x',y')]\rangle.$$ Actually, the fact that $J_{s} \propto \cos\phi_L$ can be used as an argument supporting the conjecture that $J_{s}\propto \rho_{L}(2k_F)$. To continue with the spin-vortex term, we rewrite the $\cos\phi_{L}(r_{0})$ as $$2\cos(\phi_L(r)/2)\cos(\phi_L(r')/2)-1,$$ with $(r+r')/2=r_0$ such that the extended vortex becomes a point localized at $r_0$ at large distance. Thus, there is a constant term and a term accounting for a spatial spread, the latter one is affected by the extra prefactor:
\begin{multline}
\langle\cos(\phi_L(r))\cos(\phi_L(r'))\rangle \propto\\
 \exp(\int dr\int dr' \delta t D(r-r'))=\exp(-\delta t |r-r'|),
\end{multline}
and from the screened-Coulomb gas picture we take that condensation energy has logarithmic corrections due to the spatial spread of vortex, the interaction with ``phonon" gas, then:
\begin{equation}
    J_{s}^\text{eff}=J^{0}\cdot\exp(-\delta t |r-r'|).
\end{equation}

This additional exponential can be incorporated as an extra phase shift inside the cosine, so the entire term reads:   
\begin{equation}\label{eq:H_sv(+L)}
H_{sv(+L)}=\int dr dr' y_s^{(0)}\cos(\phi_s + i \delta t |r-r'|),
\end{equation}
where there is a shift $i \delta t |r-r'|$ in the argument of the cosine. This is precisely the same term that has been treated in Horowitz\cite{Horovitz-semin}, where it was shown that this term leads to e.g. an extra rescaling of the $K_s$ TLL parameter in the equation for $\dot{y}_s$, namely
\begin{equation}
\dot{y}=y(2-K)\Rightarrow \dot{y}=y\left(2-\frac{K}{\sqrt{1-\mathfrak{d}^2}}\right),    
\end{equation}
where $\mathfrak{d}=i \delta t$. This rescaling is a result of an integral $$\int dx \cos(\delta x)K_0(x)$$ which is the Fourier transform of a modified Bessel function of the 2nd kind. This factor multiplies the TLL propagator at a given energy shell, very much like (for the second order terms) the $f_{1,2}$ functions defined in that work and generalized here in Eq.\eqref{eq:f1-def} (those are 2nd order terms, so they also have the angular $J_0(x)$ part). 

From Ref.~\cite{Horovitz-semin}, we know that the presence of the phase shift $\mathfrak{d}|r-r'|$ in Eq.~\eqref{eq:H_sv(+L)}, which is equivalent to incommensurability, modifies the RG flow for $\dot{y}_\text{LR}$: $-K\rightarrow -K/\sqrt{1-\mathfrak{d}^2}$. Curiously, when $\mathfrak{d}=i \delta t$ i.e. it is imaginary, its influence is opposite than usual because $1/\sqrt{1-(i \delta t)^2}$ is now a decreasing function of $\delta t$. This means that the term $\cos(\phi_s)$ is becoming more relevant, i.e. spin vortex fugacity is increasing in the presence of mobile orbital vortices. The imaginary character of the extra shift is a manifestation of the fact that momentum non-conserving processes in the orbital sector modify condensation energy in the spin sector. Overall, this can be thought as another version of order stabilized by a dissipation.

The situation is different for long range interactions, the $H_\text{LR}$. The interaction is mediated by the $\theta_L$ field fluctuation, so it would seem that an anomalous propagator of the $\phi_L$ term would not affect this term, although this is a quite counterintuitive result.
However, for the $H_{LR}$, in each RG step not only fields but also the vertex, which is proportional to $y_\text{LR}|r-r'|^{-\alpha}$, is affected. In a mean field approximation $\tilde{y}_\text{LR}(r-r')=\langle y_\text{LR}\cos\theta_L(r)\cos\theta_L(r')\rangle_{S_{Ltot}}$ -- the parameter of the spin long-range term is determined by averaged propagator of the orbital term. As usual in RG, we split the short (``$<$") and longer (``$>$") distance physics, \emph{both} in the spin and orbital sectors, and average the short distance action, namely:
\begin{multline}\label{eq:disord-sLR}
S_{LR(+L)}=\\ \int D\phi_{s}^{>}D\phi_{L}^{>}\bigg\{ S^{>}[\phi_{s}^{>};\phi_{L}^{>}] + 
\int dr dr'\langle \tilde{y}_\text{LR}(r-r')\rangle_{S_{L(+s)tot}}^{<}\\ \times
\langle\cos(\theta_s(r-r'))\rangle_{S_{s}}^{<}\bigg\},
\end{multline} 
where $\tilde{y}_\text{LR}(r-r')=y_\text{LR}\cos(\theta_L)$ is from the orbital degree of freedom, it contains the vertex and the propagator. The cosine in Eq.~\eqref{eq:disord-sLR} is non-local so it has to be evaluated more carefully. We use the fact that the orbital action is a sum of TLL and vortex parts $S_{Ltot}=S_{L-TLL}+S_{L(+s)vor}$, hence,
\begin{align}
   &\langle  y_\text{LR}\cos\theta_L(\vec{r})\cos\theta_L(\vec{r}')\rangle_{S_{L(+s)tot}}=\\
&\langle\cos\theta_L(\vec{r})\cos\theta_L(\vec{r}')\rangle_{S_{L-TLL}}
\langle y_\text{LR}(r\rightarrow r') \rangle_{S_{L(+s)vort}}, \notag
\end{align}
where we included the fact that correlators of $\theta_{L}$ field do not contain the vortex factor, so these factors, together with the cross-correlation with the spin-vortex sector enter only the vertex $y_\text{LR}$. The first term is a spatial average over $\theta_L$ and produces 
a known propagator $G_{L}(\vec{r}-\vec{r}')$ of the bare field $\theta_{L}$, the power law $|\vec{r}-\vec{r}'|^{-\alpha}$ in the Hamiltonian. The TLL part does not contain neither the vertex correction (by virtue of Dzyaloshinskii-Larkin theorem~\cite{DLtheo}) nor the cross-correlation with the spin sector. The vertex correction to $y_\text{LR}$ comes from an average of the total vortex part: since from Ward identity the vortex $y_\text{LR}=\Gamma_{Ls}(x=0) \sim G_{Ls}^{-1}$ and $G_{Ls}$ is a co-propagator of the spin and orbital vortex fluctuation. Here, the orbital part that contains $\langle\exp(\phi_L(r)\phi_L(r'))\rangle$ and the spin propagator of the vortices are contributing to the modified Bessel function of the second kind. We find that at each RG step there is a factor,
\begin{equation}
    1/\langle y_\text{LR} \rangle_{S_{vort}[l]}=\int_{0}^{l} dx \exp(-\delta (x)) K_0(x),
\end{equation}
with $x = |r-r'|$,  which, from the Laplace transform of a modified Bessel function of the second kind evaluates to:
\begin{equation}
    \langle y_\text{LR} \rangle_{S_{vort}[l]}=\sqrt{1-(i \delta t l )^2},
\end{equation}
with again the same imaginary factor $\mathfrak{d}=i \delta t$.

Overall the short distance average of non-local interaction term produces:
\begin{equation}
\langle\cos\tilde{\theta}_L\rangle_{S_{L(+s)tot^<}^<}=\sqrt{1-(i \delta t)^2}G_{L}(\vec{r}-\vec{r}').
\end{equation}
Upon substitution of this orbital average in Eq.\eqref{eq:disord-sLR} we obtain the usual cosine term but with a re-scaled parameter in the spin sector: $$-\frac{1}{K_s} \rightarrow -\frac{\sqrt{1-\mathfrak{d}^2}}{K_s}.$$ This enters as the last term in the last equation in the set of RG equations Eq.\eqref{eq:L-RG}. Thus, we observe that in the presence of mobile orbital vortices the \emph{spin}-LRO $\cos\theta_s$ terms are becoming less relevant. This is what we expect for the ambient pressure phase in FeSe where there is a small density of mobile bounded vortex-antivortex pairs, increasing as we approach $T_{KT}$.

\subsection{Generalized incommensurability for the vortex crystal phase}\label{sec:vortex-crystal}


\emph{Strong coupling of spin and orbital sectors.} 
In the previous section, we showed that rare and random distribution of $\vec{\eta}(x,y)$ increases the number of orbital vortices. In the strong coupling limit we need to understand what happens when the vortices get close enough such that their overlap: that is, when vortex-vortex interaction matters. This is the case ii) defined in the end of Sec.\ref{sec:UVmicro}.

Our system is such that whenever there is an orbital-vortex, there is simultaneously a preference for the FM phase in the spin system. More precisely,  at finite temperature there is a mixed state with a higher probability of the FM phase, $p_\text{FM}$. The $p_\text{FM}$ probability increases towards the center of the vortex, and decreases as we move away from it. As $p_\text{FM}(r)$ increases also the expectation value $\langle V_{LS}(r) \rangle$ increases. In what follows, we approximate this energy landscape with a set of linear potentials. In fact, a set of linear potentials is the simplest form that captures presence of space dependent $\langle V_{LS}(r) \rangle \propto \delta t p_\text{FM}(r)$, thus a generalized form of a background potential $V(r)\equiv \langle V_{LS}(r) \rangle$ where an average is taken over any slow-motion of vortexes. 

The presence of a background potential $V(r)$, results in an additional term in the Hamiltonian which has a form of a generalized ``chemical potential". It is known that such terms can be absorbed by an appropriate shift of the bosonic field:
\begin{equation}\label{eq:field-shift}
\tilde{\phi}_L(\vec{r}) \rightarrow \phi_L(\vec{r})+\int_{-\infty}^{\vec{r}} d^2r V(r).
\end{equation} 
So far this procedure has been applied to either a constant potential, for doped Mott insulator, and for a delta-like potential, for forward scattering disorder.    
A generalization merging these two procedures for coupled modes was presented in Sec.~\ref{sec:vort-disord}. Here, we need to tackle the situation of an inhomogeneous potential $V(r)$. Instead of a constant chemical potential, we apply the reasoning, Eq.\eqref{eq:field-shift}, for an (infinite) set of linear potentials which constitute an array of orbital vortices. To obtain the total potential we need to add contribution from neighboring vortices $$ V(r)\approx a r - b(r-r_1) - b(r-r_{-1})+\dots,$$ where $a,b$ are constants. We see that we obtain a linear potential in the nearest vicinity of a chosen point, namely $V(r)=(a-b)r$. 

The term in the Hamiltonian, that results from the Eq.\eqref{eq:field-shift} transformation, reads:
\begin{multline}
H_{LS(+s)}=\int d^2r ~y_L \cos(\tilde{\phi}_L(\vec{r}))=\\ \int d^2r ~y_L \cos(\phi_L(\vec{r})+ A(l)\tilde{r}^2).
\end{multline}
This term has a slightly different form than those considered previously because incommensurability induces a shift $\int dr V(r) \approx A(l)\tilde{r}^2$ which scales with a higher power of $r$ (and not linearly with $r$ as for doped Mott insulators). Here $\tilde{r}^2=x^2+\zeta y^2$ is an elliptical surface that encodes the orientational order of quantum wells ($\zeta$ is the eccentricity, that can be either along $y$, like above, or $x$ directions).  However, just like in Horowiz et al\cite{Horovitz-semin}, one expands the cosine and performs the angular integral. This results in the inclusion of extra integral coefficients in each RG equation to account for the shift. For instance the RG equation for compressibility becomes 
\begin{equation}
    \frac{dK_{\nu}}{dl}\equiv\dot{K_{\nu}}=y_{\nu}^2 f_1(l),
\end{equation}
that is the standard KT equation multiplied by a function $f_1(l)$. The functions $f_i$ contain an integral analogous to the one defined in Eq.\eqref{eq:RG-basic}. These integrals that need to be solved read,
\begin{align}
f_1(l)&= \int dr r^3 K_{1}(m r) J_{0}(A(l) r^2), \label{eq:f1-def}\\
f_2(l)&= \int dr r^3 K_{1}(m r) J_{2}(A(l) r^3), \label{eq:f2-def}
\end{align}
where $J_i$ and $K_i$ are Bessel function of the first kind and the modified second kind, respectively. The $K_{1}(m r)$ is a correlation function of the vortices  in the massive phase (with the Ising-"mass" $m\equiv y_{L}$), and Bessel functions $J_{0,2}(A(l) r^2)$ come as a result of integrating out the angular variable for monopole and quadrupole configurations, respectively. 

Recent advances in the theory of Mejer functions\cite{STOYANOV1994533,Martin_2008}, enabled us to find an explicit analytic form for these integrals:
\begin{equation}\label{eq:f1-res}
f_1(l)= c_1 Re\left[G_{2,4}^{4,1}\left(\varkappa\left|
\begin{array}{c}
 \frac{1}{2},\frac{3}{2} \\
 \frac{3}{4},\frac{5}{4},\frac{5}{4},\frac{7}{4} \\
\end{array}
\right.\right)\right],
\end{equation}
\begin{equation}\label{eq:f2-res}
f_2(l)= c_2 Im\left[\frac{G_{2,6}^{6,1}\left(\varkappa\left|
\begin{array}{c}
 -\frac{5}{6},\frac{7}{6} \\
 -\frac{1}{6},\frac{1}{6},\frac{1}{6},\frac{1}{2},\frac{1}{2},\frac{5}{6} \\
\end{array}
\right.\right)}{\varkappa^{-5/6}}\right],
\end{equation}
where $\varkappa=(y_L/A(l))^2$ contains the $l$ dependence, to be precise (as we show later in Eq.\ref{eq:varkapp-explicit}) the linear $l$ dependence but only when $\zeta\neq 0$. 
The $c_1,c_2$ are constants that are chosen in a way to ensure that $f_1=1$ when $\zeta=0$, that is in the absence of vortex crystal the RG flow is the KT flow. 

\begin{figure}[h!]
 \centering
 \includegraphics [width=0.49\textwidth]{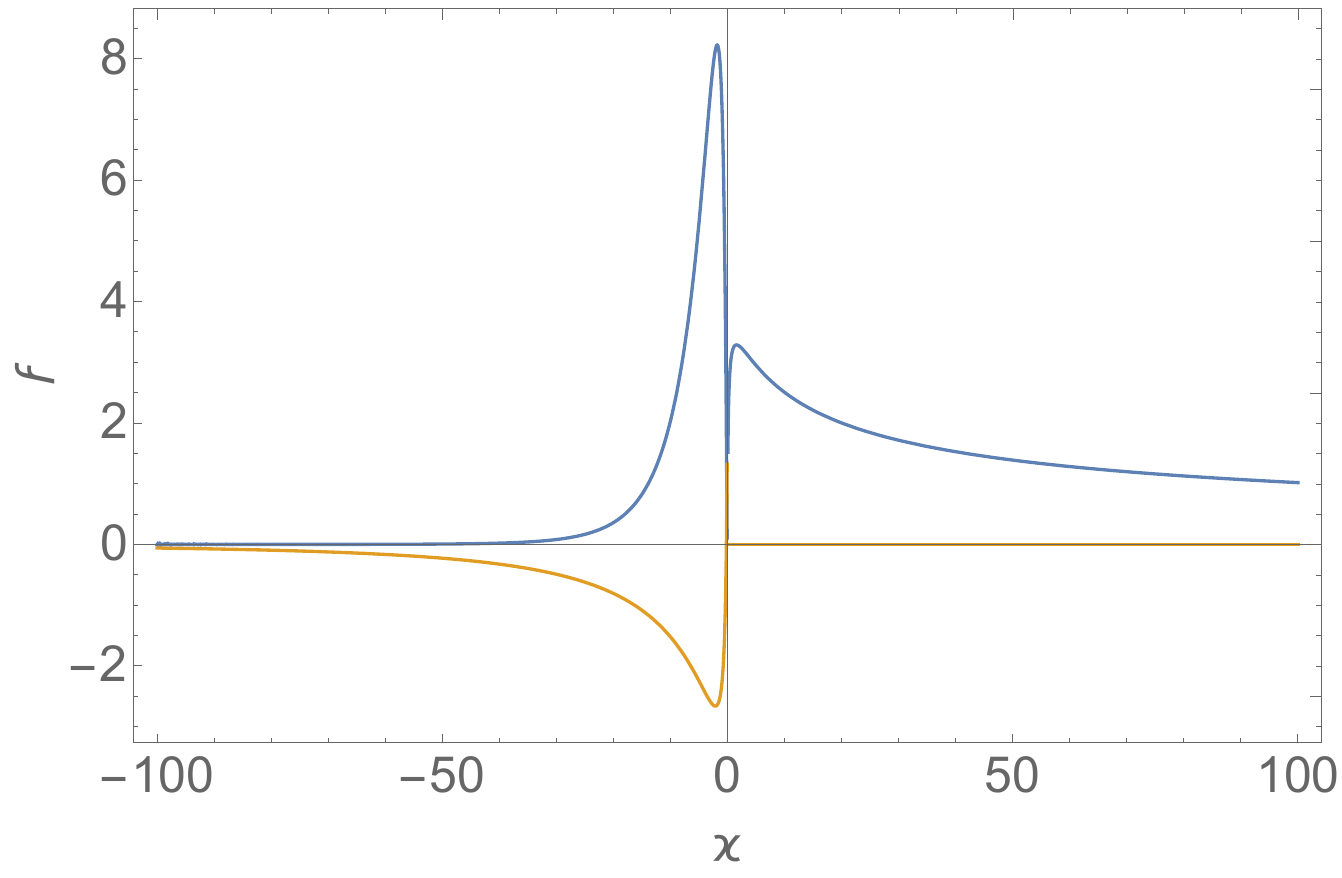}
 \caption{Extra functions $f_1$ (blue line) [Eq,~\eqref{eq:f1-res}] and $f_2$ (orange line) [Eq.\eqref{eq:f2-res}], that multiply the RG equations, shown as a function of $\varkappa$. We see that for sufficiently large positive $\varkappa$ we have $f_1=1$ and $f_2=0$, i.e. the KT flow, but once finite $\eta$ sets-in it will push $\varkappa$ to smaller values leading to a non-canonical RG flow.}
 \label{fig:integrals}
\end{figure}

We now need to connect the $A(l)$ with parameters of the theory. The coefficient $A(l)\approx (a-b)$ comes from the vortex energy landscape, which consists out of two forces in a Coulomb gas, as defined in a seminal paper by Minhagen\cite{Minhagen-rmp}:
\begin{enumerate}[I)]
\item the bare interaction, which in our case is defined by the background field $V_0(r)$, so that it is proportional to a slope $A$ for dilute vortexes which in turn is proportional to $y_L$,
\item the vortex-vortex correlation energy with its direct and orientational parts which have been expressed as $\pi/K_L\int dr' r' \chi_{v}(r')$ and $\pi/K_L\int dr' r' Ln(r'/r_v) \chi_{v}(r')$,  
\end{enumerate}
where $r'$ are positions of the dipoles in the gas.

Minhagen states~\cite{Minhagen-rmp} that including only the bare interaction reproduces the KT RG ordinary differential equation, while the vortex crystal phase arises when the vortex-vortex correlation is included. Indeed, since the argument of Eq.\eqref{eq:f1-res} depends on $y_L/V$, if we take only the first term, $V(r)/r=V_0=y_L$, then $\varkappa=1$ and thus (see Fig.\ref{fig:integrals}) the flow with constant $f_1=1, f_2=0$, i.e. unaffected KT flow. The $m^2$ can be interpreted as an inverse full (sine-Gordon) propagator minus inverse free TLL propagator $m^2=G^{-1}-G_{TLL}^{-1}$, i.e. the bare vertex function $\Gamma_0$. This observation enables us to make a straightforward generalization: to include the interaction with the surrounding environment (namely interaction with other vortices) we use an expression derived in Ref.\cite{Martin-Loss-WB} for the inverse propagator $\Gamma=\Gamma_0+\Gamma_0^4 G_\text{eff}^2$, where the second term comes from integrating out the auxiliary mode. To truncate an infinite series of equations, we assume that the dynamically generated mass term in  $G_\text{eff}$ is equal to a bubble $G_\text{eff}^{-1}=G_{TLL}^{-1}+ G_{vort}^{-1}$. This is equivalent to the ladder approximation. We then conclude that:
\begin{equation}\label{eq:varkapp-explicit}
    \varkappa=\frac{y_L^2}{\Gamma^2}=\frac{1}{1+y_L^2 G_{vort}^2 }\approx 1 - y_L^2 G_{vort}^2,
\end{equation}
where the last approximation holds in a physically relevant range of nearly unitary scattering $\Gamma\approx 1$.  We next need to evaluate $G_{vort}$ which is equivalent to the generalized force of Minhagen in point II above. The calculation is given in the App.B1b. Ultimately, we obtain the following formula for the generalized vortex-vortex ``force",
\begin{equation}\label{eq:Al-res}
 \Gamma(l)=y_L+y_L^2\frac{2 (1+Ln[1/(2\pi^2y_L)])}{(\tilde{K}^c_L-K)-y_L}+\pi\zeta l,
\end{equation}
where the vortex lattice misfit depends on the lattice distortion $\zeta$ times the characteristic scale of RG. The anomalous RG flow, the $l$-dependence of $\Gamma(l)$, begins only when $\zeta\neq 0$ and, from the first two terms in Eq.\eqref{eq:Al-res} (see Fig.~\ref{fig:integrals}) we see that this is possible only in the vicinity of KT separatrix, for a sufficiently large $y_{L}$ when $\varkappa(l)$ becomes small. In order to estimate when  the $\Gamma(l=0)$ crosses value $\Gamma(l)=0$ (on the separatrix when $y_L=K^c_L-K$) we need to solve the following equation: 
\begin{equation}
    1+y_L\frac{2 (1+Ln[1/(2\pi^2y_L)])}{(\tilde{K}^c_L-K)-y_L}=0.
\end{equation}
The latter is solved by $y_L^*=0.056$, a value quite close to Minhagen's prediction $y_L^*=0.054$.

The idea that there is an extra renormalizable parameter that determines the ``running variable" of the RG flow is in fact in close analogy to what has been done in doped Mott insulator (under constant chemical potential condition). Here, we have generalized it to the case of an inhomogeneous potential that mimics the vortex-vortex interactions.

\begin{figure}[h!]
 \centering
 \includegraphics [width=0.49\textwidth]{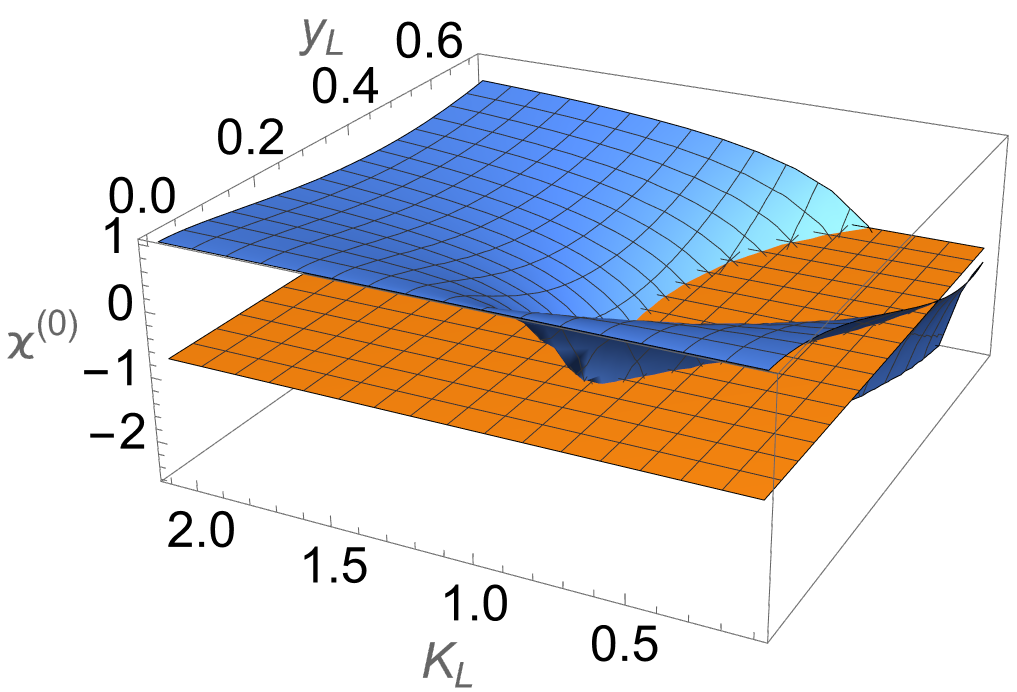}
 \caption{The blue surface shows (for chosen values of $(y_L,K_L)$) an initial value of the vertex function $\varkappa^{(0)}$(blue surface), that multiplies the RG time $l$, as given by Eq.\eqref{eq:Al-res}. By comparing with functions shown in Fig.~\ref{fig:integrals} (yellow plane indicates the value $\varkappa_{max}$ for which the function $|f_2(\varkappa)|$ has a maximum) we observe that strong deviations from canonical KT flow will appear when the blue surface lies below the orange plane. This happens only in the vicinity of the KT RG's separatrix, where indeed the vortex crystal phase is expected to emerge.}
 \label{fig:vertex}
\end{figure}
 
\section{Renormalization group}\label{sec:RG}

We are now in a position to write down the RG equations. For the sine-Gordon model, the RG equations were derived by Kosterlitz and Thouless for two phases: either with the fugacity of vortices increasing or decreasing in the infinite size limit. The generalized RG equations should capture a third phase, the vortex crystal, that is present in the vicinity of the separatrix---i.e. The line in between the two phases. Far away from the separatrix, the flow should be the KT flow. Moreover, in the vortex crystal phase, the flow must be horizontal (constant fugacity) and must gradually slowing down as we approach the new separatrix. We incorporate all effects discussed in the previous Section within a perturbative RG approach to study if the formation of the vortex crystal can be captured.

\paragraph{The orbital sector and spin-long range interactions.} The set of differential equations that describes the flow, the the Callan–Symanzik equations of the system, reads:
\begin{align}\label{eq:L-RG}
\dot{y}_L(l)&=2y_L\left(1- \frac{K_L}{\sqrt{1-\zeta^2}}\right),\\
\dot{K}_{L}(l)&=-y_L^2 f_1(l),\\
\dot{\zeta}(l)&= y_{L}^2K_L^2\left(f_2(l)-\zeta f_1(l)\right),\\
\dot{y}_\text{LR}(l)&=\notag \\ &y_\text{LR}\left(4-\alpha(1-\zeta)-\frac{\zeta}{8}-\frac{\sqrt{1-(i \delta t)^2}}{2K_s}\right), \label{eq:L-RG_last}
\end{align}
where in Eq.~\eqref{eq:L-RG_last}, we took into account the propagation of $G_L$ through ordered zones as well as the rescaling of $K_s$ in the presence of $\theta_L$ disorder. The $G_L$ propagation makes $y_\text{LR}$ more relevant ($\alpha$ is effectively smaller), while the rescaling of $K_s$ (i.e. increasing numerator in the last term) makes $y_\text{LR}$ less relevant. The competition of these two phenomena,  described in Sec.~\ref{sec:vortex-crystal} and Ref.~\ref{sec:vort-disord} respectively, clearly manifests in this equation. 

We thus see that the way to establish long-range order in the spin sector is through $\zeta\neq 0$. The non-zero $\zeta$ also affects the first equation: instead of simple KT flow, now the $K_L$ has to be re-scaled and depending on $\zeta$ value the separatrix between relevant and irrelevant $y_L$ is shifting. The system may start on irrelevant side, but later continue on the relevant side with a horizontal flow that corresponds to a constant fugacity in the orbital sector. The equation for $\dot{\zeta}$, taken together with Eq.\eqref{eq:Al-res}, is the one that determines whether or not the system will fall into the intermediate VC phase. As it can be seen in Fig.~\ref{fig:vertex}, usually the $\varkappa \approx -1$ and then from Fig.~\ref{fig:integrals} one deduces that there is a substantial, positive $f_1$ which (due to second term) suppresses any non-zero $\zeta$. However for large enough second term in Eq.\eqref{eq:Al-res}, i.e. for large enough $y_L$, the $\varkappa>0$ where $f_2\neq 0$. Then the first term in RG equation for $\dot{\zeta}$ appears, inducing growth of $|\zeta|$. From last term in Eq.\eqref{eq:Al-res} (the $l$-dependent) we see that $\varkappa$ quickly drops below zero again and then the term $\sim f_1$ suppresses the flow of $\zeta$ again. But now the finite value of $\zeta$ sets in, and the orbital sector flow is modified, deviating from the simple KT picture.


\begin{figure}[h]
    \centering
    a)\includegraphics[width=0.42\textwidth]{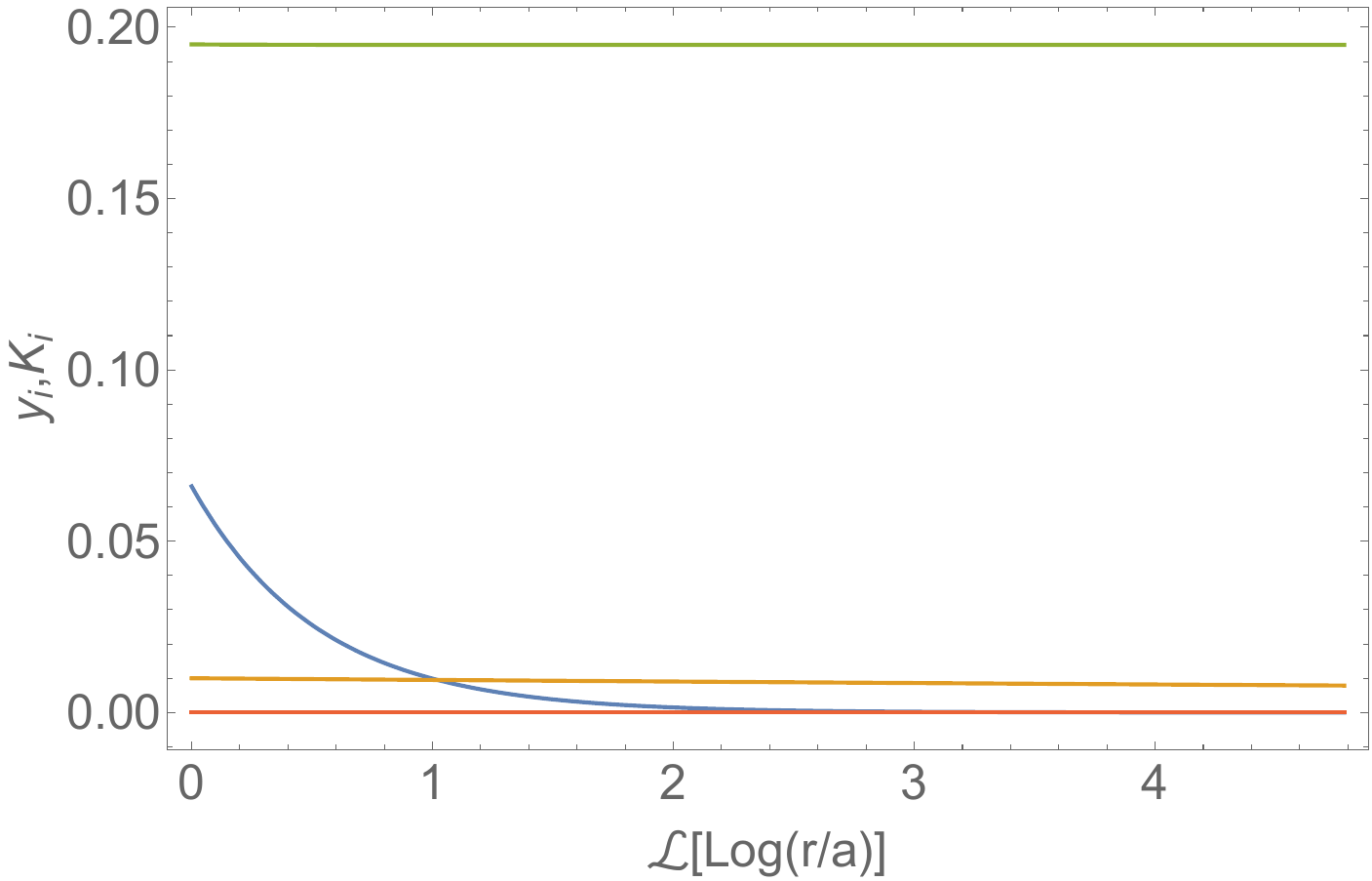}
    b)\includegraphics[width=0.42\textwidth]{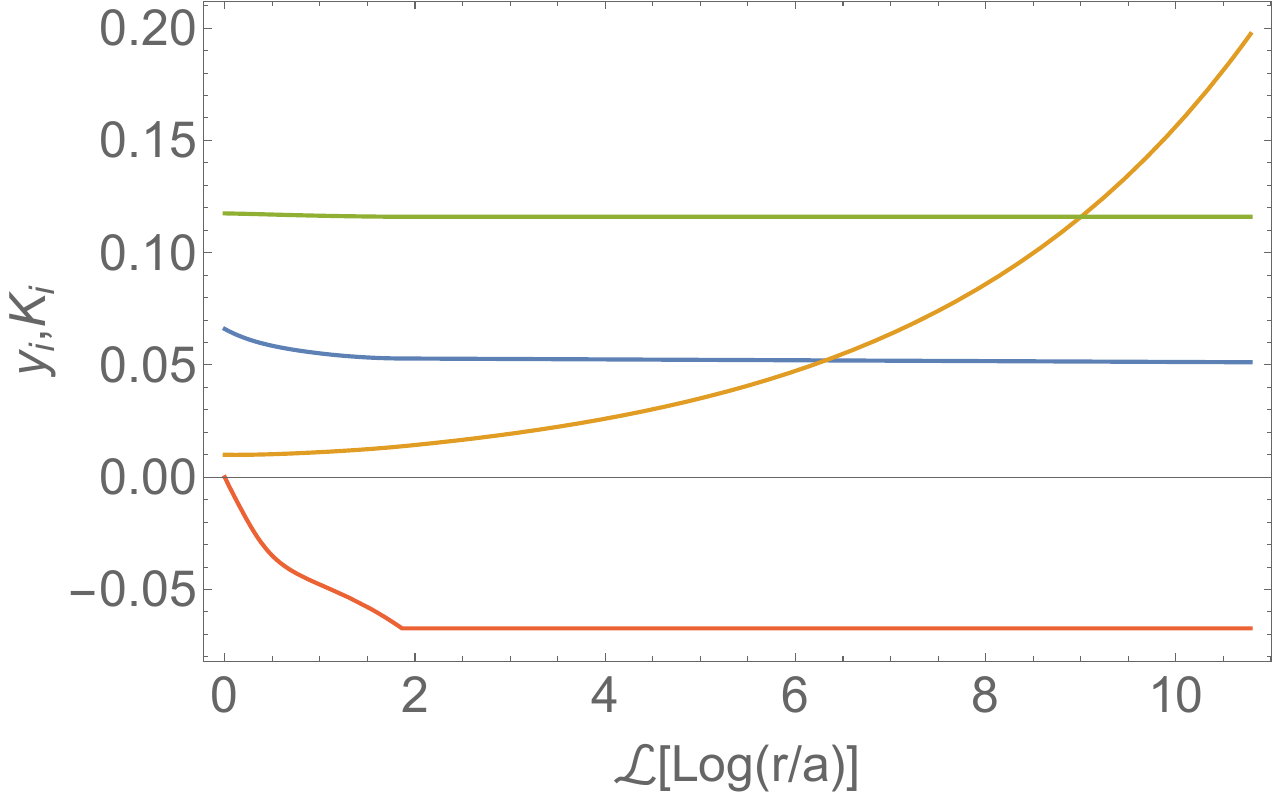}
    c)\includegraphics[width=0.42\textwidth]{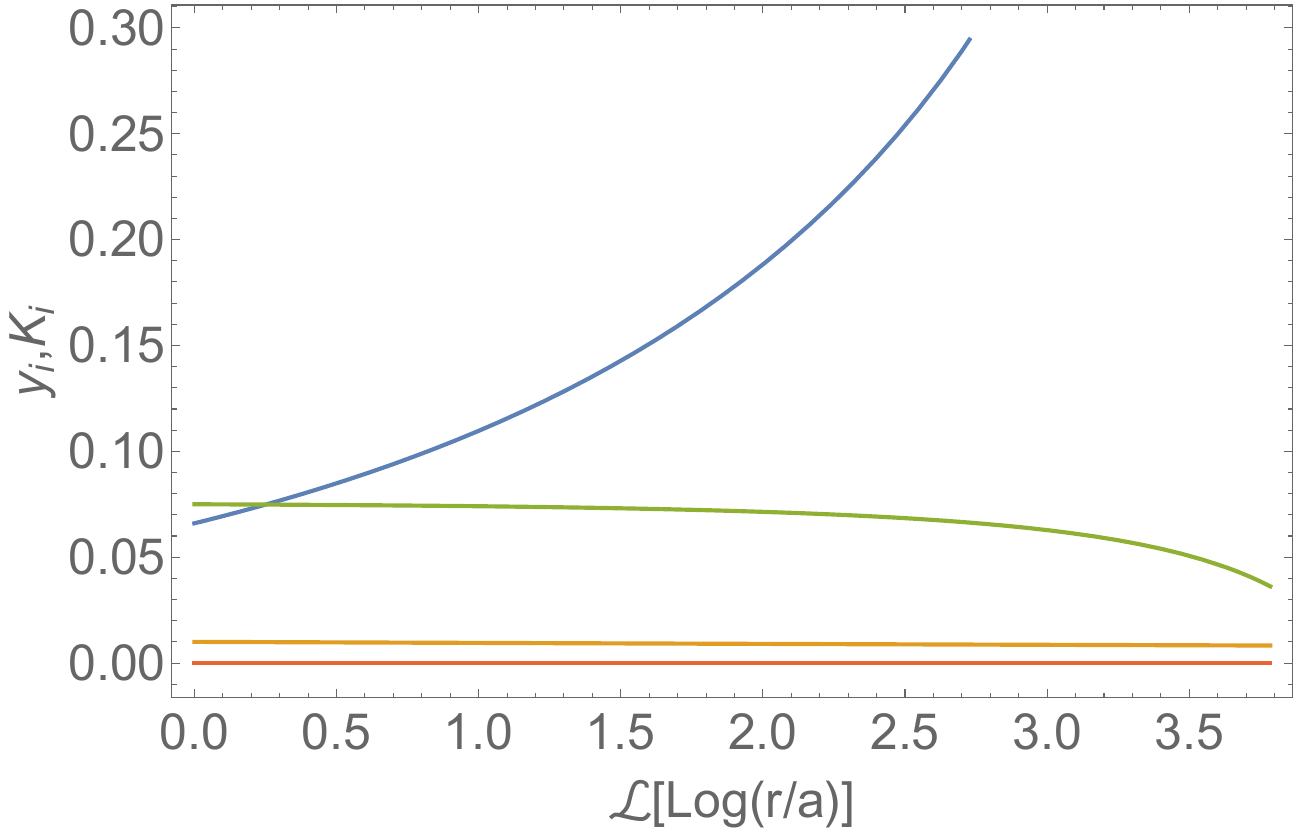}
    \caption{Three flows possible in our system. Upper and bottom panels: the standard KT irrelevant (a) and relevant (c) flows, where the fugacity $y_L$ (blue line) goes either to zero or infinity, respectively. Mid panel (b): the vortex-crystal phase where the fugacity stays constant, the orientational order parameter $\zeta$ (red line) becomes finite and the spin sector long-range order (yellow) sets in. The difference between the three cases stems from an initial ratio of $y_{L}$ (blue) and compressibility $K_{L}$ (green line).}\label{fig:RGFlows}
\end{figure} 


From Fig.~\ref{fig:RGFlows} we immediately notice how the flow is affected by the generalized incommensurability. For small $y_L$ (left panel) the $\varkappa$ stays close to one and $f_1\rightarrow 1$ while $f_2 \rightarrow 0$. Thus equation for $K_L$ is unaffected and we have a standard RG flow as expected with irrelevant $y_L$. For large $y_L>K_L^c$ (right panel) the $\varkappa$ goes to infinity and again $f_1\rightarrow 1$ while $f_2 \rightarrow 0$. Thus equation for $K_L$ is still unaffected and we have a standard RG flow  but this time with relevant $y_L$. When $y_L\approx K_L$ then equation for $K_L$ start to deviate from the standard KT and the $\zeta$ variable becomes finite. Once $\zeta \neq 0$ the functions $f_{1,2}$ start to flow as a function of the variable $l$. This will ultimately, for $l \rightarrow \infty$, gradually suppress flow in orbital sector, locking it inside the crystalline VC phase. The critical slow down of the flow is due to both the mean-field energy landscape (thus values of $f_{1,2}$) and the vertex-vertex interactions [encoded in the propagator that leads to Eq.~\eqref{eq:Al-res}]. Such slow-down of dynamics with increasing characteristic length-scale is reminiscent of a glassy phase transition.

\paragraph{The spin sector} For the spin sector we have additional equations:
\begin{align}\label{eq:S-RG}
\dot{y}_{\text{s}}(l)&= 2y_{\text{s}}\left(1-\frac{K_s}{\sqrt{1-(i \delta t)^2}}\right),\\
\dot{K}_{s}(l)&=-4y_s^2+y_\text{LR}^2.    
\end{align}
These equations do not influence much the RG flow in the bulk of the system because of the smallness of the spin fugacity---at least in the case of FeSe which serves as a potential platform of our model. Further, the presence of $y_\text{LR}$ increases $K_s$ further reducing probability of spin vortices, i.e. the fugacity $y_s$. 

However, the behavior may be different inside the core of orbital vortex (or in the vicinity of it) where $J$ is small and so the spin fugacity $y_s$ is large to begin with. We see that the randomness in orbital sector does increase even further the spin fugacity in this region. Ultimately, we reach the regime described in the Sec.~\ref{sec:vortex-crystal}, but with the role of orbital and spin-sector swapped. This will result in a periodic arrangement of spin vortices around the location of the orbital vortex. Thus the program of reaching the self-consistence proposed in Sec.~\ref{sec:UVmicro} has been achieved. 

\paragraph{Discussion} In the work of Orignac et al, Ref.\onlinecite{Orignac-biquad} an additional spin-orbit term, due to oscillatory parts of respective density terms, was included. The term reads:
\begin{equation}
    H_{stag}=J_{LS} \int d^2r \sin{(\phi_s)}\sin{(\phi_L)}.
\end{equation}
This term can appear only when the two terms $n_s^{stag}(\vec{r})=\sin{(\phi_s(x))}\exp(-i \bar{a} J(\vec{r})\cdot \vec{r}) $ and $n_L^{stag}(\vec{r})=\sin{(\phi_L(x))}\exp(-i \bar{a} J_L(\vec{r}) \cdot \vec{r}) $ cancel their oscillatory parts, that is when $J_L(\vec{r})\approx J(\vec{r})$, or more generally (to capture higher order harmonics) $J_L(\vec{r})\approx n J(\vec{r}), n \text{ is odd}$. 

When it happens then $J_{LS}$ will modify RG equations non-locally, i.e. in the areas where spin and orbital vortices co-exist. The modified RG equations read:
\begin{align}\label{eq:LS-RG}
\dot{y}_{\text{LS}}(l)&= y_{\text{LS}}(2-K_{L}-K_{s})-(y_s +y_L) y_{\text{LS}},\\
\dot{y}_{\text{s}}(l)&= 2y_{\text{s}}\left(1-\frac{K_{s}}{\sqrt{1-(i \delta t)^2}}\right)- y^2_{\text{LS}},\\
\dot{y}_L(l)&=2y_L\left(1- \frac{K_L}{\sqrt{1-\zeta^2}}\right)- y^2_{\text{LS}},\\
\dot{K}_{L}(l)&=-\left(y_L^2+\frac{y^2_{\text{LS}}}{4}\right)f_1(l),
\end{align}
where the first equation is the RG equation of the LS additional term itself.

In the limit $y_i \rightarrow 0$ the canonical phase diagram (based on the anomalous dimensions only) of the model that includes $y_{LS}$ was given in Ref.\cite{Orignac-biquad}. Clearly the model is symmetric with respect to exchange $g_s\leftrightarrow g_L$ and $K_s\leftrightarrow K_L$, so without loss of generality we can focus here only on the regime that is of our interest, namely small spin-vortex and large orbital-vortex fugacity in the bulk (close to each orbital-vortex core we will have enlarged spin-vortex fugacity). From the above equations, Eq.\ref{eq:LS-RG}, we deduce that finite, substantial $y_{LS}$ in the vicinity of orbital-vortex core (where the commensurability criterion can be met) limits the growth of other fugacities. Furthermore, it can speed up the variation of $K_L(l)$, but the limiting factor $f_1(l)$ is still there (one still needs to compute an average of $\langle\theta_L(r)\theta_L(r')\rangle$ in the vicinity of the vortex). The limiting factor induces the RG slow-down, the glassy behavior but whether it still works can be checked only numerically which we did for $y_{LS}<y_L$. On the other hand the limit of large (relevant) $y_{LS}$ can be inferred from results of Ref.\cite{Orignac-biquad} where it was shown that the phase with irrelevant $y_L$ (i.e. under the separatrix) can then be described as intertwined spin and orbital vortices, where any dislocations can lead to a glassy phase. This is not in contradiction with our proposal of orbital vortex crystal, it is worth reminding that the spin long-range order $y_\text{LR}$ was not considered in Ref.\cite{Orignac-biquad}, but a question whether there is a smooth cross-over between the two pictures is open for further studies.

\section{Measurable quantities}\label{sec:meas}

The model proposed above allowed us to write down the RG equations, and ultimately obtain the phase diagram for the strongly coupled spin-orbital system. A relevant question at this point is whether the proposed model can be in any way verified experimentally. One of the hallmarks of the model is that any quantity that contains the average of orbital fluctuations' field, the $\langle\phi_L(x,y)\phi_L(0,0)\rangle$, acquires an exponential decay. We thus need to identify some observables containing the average of orbital fluctuations' field.

A possible situation that allows to probe for the orbital sector, 
is when an electron, in its motion through the system, becomes dressed with orbital fluctuations. To model this situation, we consider the single-particle Hamiltonian [Eq.~\eqref{Eq:H1p}], the Hamiltonian for the orbital collective modes [Eq.~\eqref{eq:orb-def1}] and add a term coupling the electron with the orbital modes, 
\begin{multline}\label{eq:HelL}
    H_{e-L}=\sum_\kk \varepsilon(\kk)c^{\dag}_\kk c_\kk + \sum_\bq W_L b_\bq^{\dag}b_\bq +\\
    g \sum_\bq \sum_\kk c^{\dag}_{\kk+\bq} c_\kk^{}(b_\bq^{} + b_{-\bq}^{\dag}).
\end{multline}
In Eq.~\ref{eq:HelL}, $g$ is the momentum-independent electron-boson (orbiton) scattering parameter and we take $W_L$, introduced in Eq.~\eqref{eq:orb-def1}, to be energy associated with the boson.

We apply to $H_{e-L}$ the Lang-Firsov unitary transformation~\cite{LFtrans} to absorb the electron-boson interaction by a unitary transformation  that ``dresses" the free electrons with the boson cloud---producing orbital-polarons. It reads:
\begin{equation}
    \tilde{c_k}=c_k e^{\frac{g}{\sqrt{\omega_0}}\sum _{\text{q$\epsilon $1st}.\text{BZ}} (b_q - b_{-q}^\dag)},
\end{equation}
where for characteristic energy of the bosons we can take the energy of orbital fluctuations $\omega_0\equiv W_L$. The expectation value of the two overlapping clouds is: 
\begin{multline}\label{eq:X-def}
    X(r-r')=\\
    \left\langle \exp \left[-\frac{g}{\sqrt{\omega_0}}\sum _{q}\left(b_{-q}^\dag - b_q \right)|_{r}-\sum _{q}\left(b_{-q}^\dag - b_q \right)|_{r'}\right]\right\rangle,
\end{multline}
where the function $X(x)$ can be interpreted as an envelope that modifies electronic wave functions. Using the relation between bosonic operators and fields (in the limit of local small fluctuations):
\begin{equation}
    \left.\theta_L(r)=i\pi\sum _{q}\left(b_{-q}^\dag - b_q \right)\right|_{r},
\end{equation}
and the Debye-Waller relation $$
\left\langle \exp \left(-i\lambda \left(\theta_L(r)-\theta_L(r')\right)\right)\right\rangle=\exp \left(-\lambda\left\langle \theta_L(r)\theta_L(r')\right\rangle^2\right),$$ where $\lambda=\frac{g}{\sqrt{\omega_0}}$ is the prefactor in the exponential of the Eq.\eqref{eq:X-def}, we can reduce the problem to a computation of the bosonic field's expectation value---precisely the expectation value that we identified as a possible hallmark of our model. In other words, the coupling $g\rightarrow \tilde{g}$ is effectively modified because the static quantity is renormalized by an extra factor: the $b_L$ boson propagator $\langle b_L^{\dag}(x,y)b_L(0,0)\rangle$.

Irrespectively of what is precisely the $b_L$ boson propagator in our theory, i.e. irrespectively of the phase in which the system is located, there will be an extra exponential factor in real space, the $\exp(-\delta t |r-r'|)$ as identified in the very beginning of Sec.~\ref{sec:vort-disord}. This is an extra factor that enters the real space hybridization of polarons $t_{rr'}=t\cdot \exp(-\delta t |r-r'|)$. To obtain the reciprocal space expression, $t_{rr´}$ can be Fourier-transformed to obtain a Lorentzian with a width $\delta t \approx 2W_L$ (see App.~\ref{app:B}):
\begin{equation}
    t_{k}\propto\mathfrak{L}_D=\frac{1}{1+\left(\frac{\omega}{2W_L}\right)^2}.
\end{equation}
This renormalization has a measurable consequence. Consider a measure of the electronic dispersion $\tilde{\varepsilon}(k)$ of our system. Since the velocity is renormalized, following e.g. Feynman path integral approach~\cite{PhysRev.97.660}, there is an energy dependence~\cite{Taylor_Heinonen_2002} from the real-part of the corresponding self-energy. The latter dependence has the form of a Lorentzian $\mathfrak{L}_{\Sigma}$,
\begin{equation}\label{eq:standard-kink}
   \tilde{\varepsilon}(k)=\varepsilon(k)\times\frac{1}{1-\left(\frac{\omega}{W_L}\right)^2}. 
\end{equation}

In our system, due to its peculiar spin-orbit coupling and the resulting disorder, we have the specific real-space $t_{rr'}$ dependence derived above, so the textbook result in Eq.\eqref{eq:standard-kink} must be convoluted with the Lorentzian from the orbital vortex disorder\footnote{we use here the linear dispersion of the $b_L$ bosons} $\mathfrak{L}_D$. A convolution of two Lorentzians produces another Lorentzian  $\mathfrak{L}_{\Sigma}\otimes\mathfrak{L}_{D}=\mathfrak{L}_{\Sigma+D}$, albeit with a complex width that shifts $\mathfrak{L}_{\Sigma}(\omega)$ singularity away from real/energy axis. Then, overall, we predict the following energy dependence of the orbital-polaron dispersion's modification,
\begin{equation}\label{eq:L-polaron}
    [\tilde{\varepsilon}(k)-\varepsilon(k)](\omega)\propto \frac{W_L^2}{\omega^2+W_L^2},
\end{equation}
where we assumed that acoustic $b_L$ bosons have much lower energies than $W_L$.

\section{Discussion}\label{sec:disc}

The model considered in this work assumes the coupling between the orbital and spin sector.  A natural question is whether it is possible to model an orbital vortex crystallization without the spin sector.
In the weak coupling limit, spin degrees of freedom add extra fluctuations into the system but their velocity difference prohibits the emergence of any bound states. The picture changes substantially when we consider the presence of quasi-static FM regions. The spin sector is what justify the   \emph{assumption} of a spatially-dependent ``chemical potential", $V(r)$, and of the long-range coherence among orbital vortices contributing to the formation of a stable energy landscape. 

An open question remains whether an orbital vortex crystallization can be sustained by one mode system. Such question is not merely theoretical, in the case of FeSe, it could determine whether the vortex crystal phase persists above a pressure of 6 GPa, where the spin sector vanishes and we are left only with the orbital sector. In fact, the  VC phase was originally predicted in a single mode sine-Gordon model~\cite{Minhagen-rmp}. In such a case all bosons have the same velocity, so \emph{Migdal} theorem does not apply and the interacting modes are not separable. Additionally, the Dzyalosinskii-Larkin theorem~\cite{DLtheo} prohibits vertex corrections to the TLL part, but vortices are free to interact. A recent work~\cite{Dupuis-nonPRG} using extended non-perturbative RG has shown that the TLL parameter $K$ does depend on an amplitude of bosonic field. While this result is promising, further research is needed to fully explore its implications for the vortex dynamics. The more complex two-modes model proposed here has enabled progress, in the sense that we found the way to obtain the VC phase, but this is because of a slower spin sector that creates the energy landscape for the orbital sector. 

Though we explicitly reference to FeSe, the model proposed here is general and may be applied to diverse (quasi-)2D systems such arrays of Josephson junctions or incipient ferroelectrics. The application to a different system only requires to change the initial parameters of the RG flow. 
For this work, the initial parameters allow to study FeSe properties around the temperature of the parent state of any ordering ($\approx 100$~K). In principle, taking the $T\rightarrow 0$ would allow, in the absence of instabilities, to reach the 2D ground state. Nevertheless, if $y_\text{LR}$ remains large at large distances, the spin system tends towards an instability and at lower temperatures the system is describe by a different Hamiltonian. 

As a final point, Eq.~\ref{eq:L-polaron} has the form of a coupling with a system characterized by an imaginary frequency. This feature highlights a connection to the well-known Hatano-Nelson model~\cite{PhysRevB.56.8651}, as we also analyze vortex-vortex interactions and independently derive a term resembling an imaginary electromagnetic potential. This connection places our findings in Sec.\ref{sec:vort-disord} within a broader context.

 \section{Conclusions}\label{sec:conc}
From a theoretical point of view, the key contribution of this work is the derivation of a RG scheme capturing the intermediate vortex crystal phase that exists at large enough fugacities. The existence of this phase was already inferred three decades ago~\cite{Minhagen-rmp} and a mean-field analysis was performed~\cite{gabay93} shedding light on the properties of this phase. However, to our knowledge the unifying RG has been missing. To formulate a unifying RG, we treated the problem of generalized incommensurability and uncover the orientational variable that drives the RG flow. The observation that the orientational variable $\zeta$ governs the switching between the minima of the cosine potential aligns our work with the recently proposed picture in Ref.~\cite{coleman-numer} for an intermediate hexatic phase.

From an experimental point of view, when considering FeSe---the study-case of this work---the most significant result is the influence of the orbital sector on the spin sector, and its consequences. Specifically, the model here presented detailed how the spin LRO and the orbital vortex crystal are inter-twinned and support each other. In 2D, fluctuations that typically destroy any form of order are suppressed, absorbed into topological orbital-vortex states that subsequently organize into a regular pattern. This stabilization allows the system to develop long-range order in the spin sector. Thus, the model proposed explains why in FeSe a proper spin order appears only at finite pressure and predicts that a nanoscopic size domain of spin LRO can appear at ambient pressure. Finally, we found that the presence of orbital fluctuations can be probed experimentally measuring the electronic dispersion.

\appendix

\section{Derivation of the model}~\label{app:A} 

The Hamiltonian of our interest, Eq.\eqref{eq:ham} can be realized in various physical realizations: two coupled distortive modes (of the quasi-2D incipient ferroelectric lattice) or 2D Josephson junctions array with pair tunneling depending on the charge state of each Cooper-box, to name a few examples.

While the formalism proposed in this paper is completely general, for concreteness we focus on FeSe, a multi-orbital quasi-2D periodic crystal where unusual coupled spin and orbital physics has been found in \emph{ab-initio} studies. From DFT Kohn-Sham (KS) calculations one obtains an effective single particle description diagonalized in a momentum space: 
\begin{equation}\label{Eq:H1p}
H_\text{1p} = \sum_{\kk}\varepsilon_{l,\kk}c^{\dag}_{l,\kk}c^{}_{l,\kk},   
\end{equation}
where $\varepsilon_{l,\kk}$ are the KS eigenenergies and $c^{\dag}_{l,\kk}, c_{l,\kk}$ the creation and annihlation operators corresponding to the KS eigenstate with spin-band index $l$ and crystal momentum $\kk$.

By moving to basis of Wigner states this kinetic part maps the Kohn-Sham results into a tight-binding Hamiltonian, 
\begin{equation}\label{eq:ferm-tight-bind}
   H_{1p} = \sum_{\xi} \epsilon_\xi \sum_{i\sigma} c^\dagger_{i\xi\sigma} c^{}_{i\xi\sigma} - t_\xi \sum_{ij,\sigma} c^\dagger_{i\xi\sigma} c^{}_{j\xi\sigma},
\end{equation}
where $i,j$ indicate the unit cell, $\sigma \in \{ \uparrow, \downarrow\}$ indicates the spin, $\xi$ indicates the band index. $\epsilon_\xi$ and $t_\xi$ are, respectively, the on-site and hopping constants for each band $\xi$. $c^\dagger_{i\xi\sigma}$ and$ c^{}_{i\xi\sigma}$ denote the creation and annihilation operator for a fermion at unit cell $i$, with spin $\sigma$ and band $\xi$. We see that the single-particle eigenenergies $\varepsilon_{l,\kk}$ can be in principle parameterized by a set of tight-biding parameters $t^{(l)}_{rr'}$.

We know that electrons do interact with each other and full model has to take into account these correlation effects. The above given, essentially mean-field description, cannot capture all interaction effects thus there should be an additional term in the complete Hamiltonian:
$$
H_{int}=\sum_{\alpha\beta\gamma\delta}\int d^d r c_{\alpha}^{\dag}c_{\beta}^{\dag}V_{int}(\vec{r};\alpha\beta\gamma\delta)c_{\gamma}c_{\delta},
$$
where $\alpha\beta\gamma\delta$ summation runs over all spin, orbital degrees of freedom. One way to capture effect of interaction is to select a decoupling channel, e.g. density-density, define a bilinear boson $b_{\nu}\propto \rho_\nu$ and then perform Hubbard-Stratonovich transformation. This is equivalent to completing a square $H_{int}\rightarrow H_{int}-D_{bos} b_{\nu}^{\dag}b_{\nu}$ where the propagator $D_{\nu}^{b}$ is proportional to inverse interaction $V_{\nu}^{-1}$. In general $\rho_{\nu}$ can be any linear combination of $\rho_{\alpha\beta}$ thus in choosing the $\nu$-th de-coupling channel we anticipate which density fluctuations are presumably important in the system. Choosing $\rho_{\nu}$ also determines $V_{int}(\vec{r},\nu)$ and in turn the dispersion of the collective modes.

From the theoretical point of view computing directly the part of interaction that has been absorbed by a given de-coupling $V_{int}(\vec{r},\nu)$ is a quite difficult task. Fortunately one can apply the opposite method: compute the dispersion of fluctuations that correspond to a given modification of the density. If the dispersion can be fit well with a cosine, then a tight binding model description for bosons, an analogue of fermionic Eq.\eqref{eq:ferm-tight-bind}, can be made. This also builds a connection with a \emph{local} density displacement.

We now aim to identify the $\rho_{\nu}$, for orbital degree of freedom $\nu=1,2(orb)$. To this end we consider the \emph{local} displacements. We assume that on the atomistic level the two relevant states are two degenerated $d_{xz},d_{yz}$ atomic Fe-orbitals (with $L_z=1$) and $d_{xy}$ atomic Fe-orbitals (with $L_z=2$), thus we may consider it either as a two-state system with an action $S_{orb}[\rho_{\nu}]$ that changes the value of $\nu'\rightarrow\nu\pm 1$ keeping it in $L_z=1,2$ manifold, or as an effective pseudo-spin $|\Tilde{L}|=1/2$ with $\rho_{\nu}^{\mathfrak{j}}=\Tilde{L}^{\mathfrak{j}}=c_{\nu}^{\dag}\hat{\nu}_{\mathfrak{j}} c_{\nu'}$ where $\hat{\nu}_{\mathfrak{j}}$ is a Pauli matrix in the two-orbital space. The second model has been studied quite intensely in the field of magnetism. One way of treating such a pseudo-spin problem is through Holstein-Primakoff transformation, with an advantage that the relevant $\Tilde{L}^{\pm}$ simply corresponds to creation/annihilation of a local boson $b_L^{\dag}(x_j),b_L(x_j)$ on the site $x_j$ of the lattice. This is the boson that de-couples interaction in the Hubbard-Stratonovich transformation. In this way we establish the connection between collective modes characteristic of itinerant system and pseudo-spins that can be also used to characterize localized carriers. In the itinerant regime the boson description is that of a screened inter-orbital plasmon, i.e. acoustic $\omega_L(q)\sim q$, while in the presence of spin-orbit coupling (and an underlying AFM phase) in the localized regime one also finds $\omega_L(q)\sim q$.

In the presence of electron-electron interactions there will be two-body contributions to total energy of the system, that depends (in the simplest case) on densities $\rho_{\alpha\beta}$. This justifies the choice of $\rho_{\nu}$, namely if energy depends on orbital configurations then fluctuations encoded by $b_L$ bosons ought to be taken into account. Let's define the bosonic operator, 
\begin{equation}
b_L^{\dag}(x,y)= \sum_{\kk} c_{l,\kk}^{\dag}(x,y)\hat{\sigma}_{ll'}^+c_{l',\kk}(x,y),
\end{equation}    
where $\hat{\sigma}_{ll'}^+$ is a Pauli matrix acting in the space of orbitals. The fact that overall energy depends on orbital occupancy, $E_{tot}[n_L]$, with $n_L\equiv b_L^{\dag}b_L^{}$, indicates that there is a corresponding non-zero term in the Hamiltonian of the system that we should account for.

\section{RG equations details}~\label{app:B}
\subsection{Orbital sector}~\label{app:B1}
\paragraph{Incommensurability in terms of Meijer G-function} In order to arrive at Eq.\eqref{eq:f2-def} we  separate the part of action for the highest energy degrees of freedom:
\begin{equation}
    S=S_{<}+S_{>},
\end{equation}
where each of the action reads:
\begin{multline}
    S=\frac{u}{2\pi K}\int d^2r [(\nabla_x \theta)^2 + (\nabla_y \theta)^2]\\
    +y\int d^2r\cos(\theta),
\end{multline}
but the shortest range fluctuations are separated:
\begin{equation}
    \theta_{>}(x,y)=\sum_{\Lambda'<|\vec{q}|\Lambda}\exp(i \vec{q}\cdot\vec{r})\theta(\vec{q}).
\end{equation}
For the quadratic part of action the procedure is straightforward, the action is separable, but the cosine need careful attention. We use the formula for the cosine of a sum of two angles, then expand the latter part to average out (see App. E in TG book for details) shortest distance fluctuations. The first order term $\propto y$ gives renormalization of $g$ while the second order term, that leads to $f_{1,2}$ functions is of the form:
\begin{multline}\label{eq:2nd-order-RG}
    \delta I =
    \int d\vec{r}_1\int d\vec{r}_1\cos(\theta_{<}(\vec{r}_1)-\theta_{<}(\vec{r}_2))\cdot\\
    [\exp{\langle(\theta_{>}(\vec{r}_1)-\theta_{>}(\vec{r}_2))^2\rangle}],
\end{multline}
that we have integrate out to obtain a given RG step. One makes a change of coordinates $(x,y)\rightarrow (r,\varphi)$ and then integrate over the angular variable $\varphi$. In case of incommensurability one has to substitute $\theta(x,y)\rightarrow\tilde{\theta}(x,y)$. In the standard case (chemical potential inducing $\cos(\delta x)$ under the integral Eq.\eqref{eq:2nd-order-RG}) there is simply $x^2 - y^2 = r^2 \cos(2\varphi)$ which upon angular variable integration leads to $J_2(\delta l)$.

Here however there is a lattice of vortices which implies that there is a skew ($\pi/2$ phase shift) between distortion $\zeta$ along x and y directions. We need to add and extra factor $\exp{i\pi/2}=i$. Moreover, assuming linear potential $V(x)$ (saw-tooth) we notice that contributions along x and y cancel at this order, hence for this \emph{relative} average we need to take the second order $V(r)\sim r^2$. This leads to Eq.\eqref{eq:f2-res} for the function $f_2$.

Another generalization in our reasoning is that we need to distort the space itself $r\rightarrow \tilde{r}$, i.e. the angle $\varphi$ now describes rotation on an ellipse instead of a circle. In our reasoning so far we assumed that direction of $\zeta$ (the eccentricity of ellipses) is fixed. This is not true, in fact we have a degree of freedom associated with the versor $\hat{\zeta}$ that is described by Ising model:
\begin{equation}\label{eq:Ising}
    H_{\hat{\zeta}}=J_{\hat{\zeta}}\sum_{j}\hat{\zeta}_{j}\hat{\zeta}_{j\pm 1},
\end{equation}
where $j$ index counts the vortices and the variable $\hat{\zeta}$ can take only two values. We see that continuous rotation invariance (the $U(1)$ symmetry) has been reduced to a discrete $Z_2$ variable, an intermediate stage before complete gaping these fluctuations. 

\paragraph{Vertex correction due to vortices} We now need to evaluate $G_{vort}$ which proceeds along the similar lines like computations in Sec.~\ref{sec:vort-disord} (i.e. propagator convoluted with exponential decay). This quantity is actually equivalent to the generalized vortex-vortex force of Minhagen if one notices that $\chi_v \equiv \langle n(r) n(0)\rangle=G_{vort}$ (the boson propagator is a two body propagator in an underlying Coulomb gas model of Minhagen).

In order to evaluate the integral that gives boson propagator we notice that the susceptibility $\chi_{v}(r')\sim y_L$ times fluctuations around minima and the distance between vortices changes with the length-scale as $r_{v}+\exp(\zeta l)$ so if we expand then we shall have a term $y_L \zeta l $ (times a constant) plus a term from the regular vortex lattice. The positions of vortices inside vortex crystal can be approximated as a cosine function. We then arrive at integral of the form $\int dr' Cos(r'/r_v) K_{1}(m_v r')$ which can be well approximated by a Fourier transform of the Bessel function of the second kind $\sim \sqrt{1+m_v^2}$. 

The mass $m_v$ that enters in the argument of Bessel function is a soliton mass of orbital breather which is known to be proportional to $y_{L}/(\tilde{K}_L^c-K_L)$ (where we expanded the $sin(nK/(\tilde{K}_L^c-K_L))$, since we are interested in the lowest energy excitation). Moreover we take into account Ising fluctuations of massive phase (so the orbital sector is not completely frozen) by shifting $\tilde{K}_L^c\rightarrow K_L^c - 1/8$, i.e. it is as if the dimensionality power counting was modified.

\subsection{Spin sector}~\label{app:B2}

\paragraph{The $\zeta$ dependence in $\dot{y}_\text{LR}$} In the main text we write an equation:
$$
\dot{y}_{LR}=y_{LR}(4-\alpha(1-\zeta)-\zeta/8-1/(2K_S)),
$$
which is similar to the one obtained in Ref.\cite{Maghrebi-CSB1D}, but with some notable differences:
\begin{enumerate}
\item the scaling dimension is $4$ instead of $3$, this is because we compute interaction with relative 2D distance variable $(r_1-r_2)$ which implies a four dimensional integral $\int dx_1 dy_1 dx_2 dy_2$, this change follows the same argument as it has been advocated in Ref.\cite{BKT-long-range} where the correspondence between 2D classical and 1D quantum systems with long-range interactions was established
\item instead of $\alpha$ we have now more complicated expression $\alpha(1-\zeta)-\zeta/8.$
\end{enumerate} 

In order to understand the second modification we need to remind ourselves that in our model $\alpha$ is not a free parameter but, when orbitons are mediating spin-spin interactions, the $\alpha$ is equal to spatial decay of orbiton propagator. Thus it emerges from averaging over $S_{LR(+L)}$. Following path integral formulation we divide the path into $N-$sections and then the overall time evolution operator is $\hat{T}=[\hat{T_1}*\hat{T_2}*...*\hat{T_N}]^{1/N}$ which in the case of the same power-law law decay in each section gives $\hat{T_i}=\hat{T_0}$ where $\hat{T_0}$ is a free particle decay. The situation becomes more complicated when vortex crystal sets-in in a fraction $\zeta$ of the sample (or it is present at a given site with probability $\zeta$ in a quantum mechanical framework). Then some of the time-evolution operators will be those of a free propagator, while some will be those of a vortex crystal. Naively, the orbiton should not decay inside VC, however in our case we have two possible orientations of distortion (x or y), just like in a twinned crystal, hence there is an Ising variable, see Eq~\ref{eq:Ising}. The additional $S_{LR(+L)}$ is now an action of the 2D Ising model. Its correlation functions decay is as usual following $1/r^{1/8}$ power-law dependence. In general it would be hard to argue that time evolution operators $T_0$ and $T_I$ commute, however here we are only averaging out the effect at a given shell of distances which limits phase space for any interference effects. We then group the time evolution operators as follows:
$$
\hat{T}=[\hat{T_0}^{N-\zeta N}*\hat{T_I}^{\zeta N}]^{1/N},
$$ 
which directly leads to an effective exponent of the spin-spin interactions' decay included in RG equation.  

\emph{Remark:} For a free theory the orbitons decay is $\alpha=4-1/(2K_L)$ and since $K_L$ is relatively small (to access the vortex crystal phase, in the vicinity of separatrix), certainly $K_L<K_S$, then exponent is reduced slightly below the critical value, because $1/K_L > 1/K_s$. This last relation may change if we change $-1/K_s \rightarrow -\sqrt{1+(i\delta t)^2}/K_s$, thus increasing the orbital motion pushes the long-range spin interactions $y_\text{LR}$ towards irrelevance.

\section{Prediction for $\delta t$}

Let us assume that there exists a finite amplitude of non-local exchange $J_{non-loc}$ that is proportional to $J_{H}$ times a coherent spread of the electronic orbital. Actually, it has been shown\cite{Held-Jpropt} that this term is related to, proportional to (for the time-reversal invariant case equal to) the pair-hopping probability which justifies our notation $\delta t$. In principle, it can be taken as a free parameter, but it ought to be also related to the dispersion of orbital fluctuations. We shall elaborate on this connection here.

As stated in the main text: if two carriers located on neighboring (more generally covalently bound) sites have parallel spins then thanks to indistinguishability principle they can be exchanged and energy can be gained. This is encoded in a finite $J_{non-loc}$. Of course, what is \emph{assumed} here is that these two carriers will stay for an infinitely long time in this bound state. Let us assume that the \emph{local} FM state is long-lived and entire time evolution is coherent. Even then the effective exchange coupling will read:
\begin{equation}
    \delta\tilde{t}= J_{non-loc}(1-p_{esc}),
\end{equation}
where $p_{esc}$ is the escape probability that in the zeroth order is equal to $(1-p_{rtn})/2$ where the $2$ in the denominator is due to neighbors on two sides and $p_{rtn}$ -- probability of return to previous orbital configuration is $W_L/J_{non-loc}$ times $W_L/J_{non-loc}$ (like an orbital super-exchange with an energy cost, the interaction, equal to $J_{non-loc}$). 

From now on the procedure can be continued recursively: the two in the denominator is there because of two potential escape paths, but each of these is reduced due to the return probability. One continues this procedure towards further and further neighbors, which results in a continued fraction:
\begin{equation}
    p_{esc}=\ContFracOp_{m=1}^n \frac{-(1-(W_L/J_{non-loc})^2)}{2},
    \label{eq:contfrac}
\end{equation}
where $\ContFracOp_{m=1}^n$ is a Gauss continued fraction symbol. Remarkably, the above given infinite expression simplifies to:
\begin{equation}
    p_{esc}=1-\sqrt{(W_L/J_{non-loc})^2},
\end{equation}
which immediately implies that:
\begin{equation}
    \delta t = J_{non-loc}\sqrt{(W_L/J_{non-loc})^2}=W_L,
\end{equation}
Thus indeed there is a simple relation between the amplitude of orbital fluctuations and the non-local tunneling. The above expression is in fact a limiting value in the case when FM domains are infinitely long-lived and exchange processes continue to propagate coherently throughout the lattice. 
In fact the largest value of $\delta t$ is $W_L$. For any larger value the two minima swap and the role of the two phases is exchanged.

\bibliography{bib}
\end{document}